\documentclass[aps, prd, showpacs, superscriptaddress, ctexart, nofootinbib, twocolumn]{revtex4}
\usepackage{amssymb,amsmath,graphicx,color,microtype}

\usepackage{color}

\begin{document}
\title{Exploring Bouncing Cosmologies with Cosmological Surveys}

\author{Yi-Fu Cai}
\email{yifucai@physics.mcgill.ca}
\affiliation{Department of Physics, McGill University, Montr\'eal, QC, H3A 2T8, Canada}

\begin{abstract}
In light of the recent observational data coming from the sky we have two significant directions in the field of theoretical cosmology recently. First, we are now able to make use of present observations, such as the Planck and BICEP2 data, to examine theoretical predictions from the standard inflationary $\Lambda$CDM which were made decades of years ago. Second, we can search for new cosmological signatures as a way to explore physics beyond the standard cosmic paradigm. In particular, a subset of early universe models admit a nonsingular bouncing solution that attempts to address the issue of the big bang singularity. These models have achieved a series of considerable developments in recent years, in particular in their perturbative frameworks, which made brand-new predictions of cosmological signatures that could be visible in current and forthcoming observations. In this article we present two representative paradigms of very early universe physics. The first is the so-called new matter (or matter-ekpyrotic) bounce scenario in which the universe starts with a matter-dominated contraction phase and transitions into an ekpyrotic phase. In the setting of this paradigm, we propose some possible mechanisms of generating a red tilt for primordial curvature perturbations and confront its general predictions with recent cosmological observations. The second is the matter-bounce inflation scenario which can be viewed as an extension of inflationary cosmology with a matter contraction before inflation. We present a class of possible model constructions and review its implications on the current CMB experiments. Last, we review the significant achievements of these paradigms beyond the inflationary $\Lambda$CDM model, which we expect to shed new light on the direction of observational cosmology in the next decade years.
\end{abstract}


\maketitle

\section{Introduction}\label{sec:intro}

The first evidence for the primordial B-mode polarization of the cosmic microwave background (CMB) has very likely been detected by the Background Imaging of Cosmic Extragalactic Polarization (BICEP2) telescope at the South Pole in March of 2014 \cite{Ade:2014xna}. Its statistical significance is claimed to be above the $5\sigma$ confidence level (C.L.). The simplest and also the most reasonable theoretical interpretation of these signals is the primordial gravitational wave which was arisen from the ripples in the spacetime in the very early universe. Obviously, this important measurement, together with another CMB observation -- the Planck data released in 2013 \cite{Ade:2013zuv}, have brought major information to the field of very early universe physics and theoretical physics in general.

The recent developments in observational cosmology has a series of significant implications on inflation models. In this picture, cosmologists believe that when the universe was very young, namely about $10^{-35}$ second after the big bang, it has undergone a period of extremely rapid expansion, known as "inflation", when its volume increased by a factor of up to $10^{80}$ in less than $10^{-32}$ second. This paradigm, as proposed in the early 1980s to understand the initial condition issue in the hot big bang cosmology \cite{Guth:1980zm, Linde:1981mu, Albrecht:1982wi} (see also \cite{Starobinsky:1980te, Fang:1980wi, Sato:1980yn} for early works), has so far become the most prevailing in describing the very early universe. Its critical prediction of a nearly scale-invariant power spectrum of primordial curvature perturbations has been confirmed to high precision by the observations of the CMB temperature anisotropies in recent years \cite{Ade:2013zuv}. Inflation also predicted a nearly scale-invariant power spectrum of primordial tensor perturbations \cite{Starobinsky:1979ty}, which can seed the B-mode polarization as detected by the BICEP2 experiment. Provided that all polarization signals were contributed by inflationary gravitational waves, one can apply the recent data to constrain inflation models.

Although a major achievement has been made, one ought to be aware of the fact that inflationary cosmology is not the unique paradigm describing the very early universe. Moreover, general inflation models have difficulty in explaining the Planck and BICEP2 data simultaneously. Additionally, we will point out in the main text that there exist several conceptual challenges that cannot be resolved in the context of inflationary cosmology. Having the recent experimental developments and theoretical challenges in mind, we are motivated to search for new proposals for the theory of the very early universe beyond the paradigm of inflationary $\Lambda$CDM cosmology. In the literature there are indeed several alternative scenarios to inflation that describe the very early moment of the universe. Namely, the ``pre big bang" scenario \cite{Gasperini:1992em} suggests that the universe may start from an initial state characterized by a very small curvature, based on the scale-factor duality \cite{Veneziano:1991ek} of string cosmology. Moreover, the ``ekpyrotic/cyclic universe" configuration \cite{Khoury:2001wf} suggests that the periodic collision of two membranes in a high-dimensional spacetime could give rise to a cyclic solution. One another interesting paradigm of very early universe is the so-called ``emergent universe" cosmology \cite{Ellis:2002we}, in which our universe emerges from a non-zero minimal length scale and experienced a sufficiently long period of quasi-Minkowski expansion to then begin the normal big bang expansion. This scenario was originally achieved in the ``string gas" cosmology in terms of a Hagedorn phase of a thermal system composed of a number of fundamental strings \cite{Brandenberger:1988aj}. Recently it was also realized by a cosmic fermion condensate model \cite{Cai:2012yf, Cai:2013rna}. In addition, from the perspective of phenomenological considerations, there is the ``matter bounce" scenario \cite{Wands:1998yp, Finelli:2001sr} that gives rise to almost scale invariant power spectra of primordial perturbations and thus can fit to observations very well. We refer to \cite{Brandenberger:2011gk} for a comprehensive review of various proposals of very early universe models.

Amongst all of these competitive configurations, we focus in this article on two phenomenological scenarios alternative to the inflationary $\Lambda$CDM model with one being the healthy nonsingular bouncing cosmology \cite{Cai:2012va} and the other the matter-bounce inflation scenario \cite{Cai:2008qb}. The idea of the bouncing cosmology appeared even much earlier than inflationary cosmology, namely, one can trace back to Tolman's work in the 1930s \cite{Tolman:1931zz}. A modern version of bouncing cosmology suggests that the expansion of the universe is preceded by an initial phase of contraction and then a non-vanishing bouncing point happened to connect the contraction and the expansion \cite{Mukhanov:1991zn}. It was recently pointed out in Ref. \cite{Cai:2007qw} that if such a bounce model describes a realistic universe within General Relativity, then motivated by the study of dark energy physics \cite{Feng:2004ad} a matter field with a so-called quintom scenario is needed that the equation-of-state parameter has to evolve below the cosmological constant boundary $w=-1$ twice. This type of bounce model was extensively studied in the literature in recent years (for example see \cite{Cai:2009zp} for a related review of Quintom cosmology and references therein for extended analyses). However, it is not easy to construct a model that can realize a nonsingular bouncing solution without theoretical pathologies since the quintom scenario is associated with the violation of the null energy condition (NEC) which is accompanied with various quantum instabilities \cite{Cline:2003gs}. Furthermore, a general challenge for bouncing cosmologies is to ensure that their contracting phases are stable against the instability to the growth of some unexpected anisotropic stress, whose associated energy density grows as $a^{-6}$. This is recognized as the famous Belinsky-Khalatnikov-Lifshitz (BKL) instability issue of any cosmological models involving a contracting phase \cite{Belinsky:1970ew}. After years of unremitting efforts, it was recently proposed in Ref. \cite{Cai:2012va} that an effective field description that combined the benefits of matter bounce and ekpyrotic scenarios can give rise to a nonsingular bouncing cosmology without pathologies through a Galileon-like Lagrangian as well as explaining the formation of the large scale structure (LSS) in our universe.

The matter-bounce inflation scenario is another promising paradigm of the very early universe in which the inflationary cosmology is generalized by introducing a matter-like contracting phase before the inflationary phase. The scenario of bounce inflation was earlier proposed in \cite{Piao:2003zm} where the universe is assumed to evolve from a phase of fast-roll contraction to the inflationary era with a nonsingular bounce in between, and hence, the power spectrum of primordial curvature perturbation becomes strongly blue at extremely large scales \cite{Liu:2013kea}. Alternatively, if one makes an analogue with the evolution of our universe, it seems more natural to consider the universe to be dominated by regular matter fields in the contracting phase. As a result, the scenario of matter-bounce inflation was implemented by a double-field quintom model in \cite{Cai:2008qb}. Within this scenario, the power spectra of primordial curvature perturbations are almost scale-invariant at both large and small length scales while some local features appear due to the phase transition \cite{Cai:2008ed}. A representative phenomenon is that the amplitude of the power spectrum of primordial curvature perturbations could experience a step feature at a critical scale, which may be applied to explain the suppression of the low $\ell$ CMB multipoles as observed by the WMAP and Planck experiments \cite{Liu:2010fm}. Moreover, the tensor spectrum of this scenario can have a step feature as well and thus, it allows for more freedom in the parameter space to fit to the CMB observations. In particular, this scenario was found to be very powerful in reconciling the tension between the Planck and BICEP2 observations when compared with the $\Lambda$CDM \cite{Xia:2014tda}. Additionally, from a top-down viewpoint this scenario can easily be embedded into some fundamental theories, for instance, the loop quantum cosmology \cite{Mielczarek:2010bh}.

The article is organized as follows. In Section \ref{sec:inflation}, we briefly list the main predictions made by the standard inflationary cosmology. While these results are roughly consistent with the present observations, we intriguingly point out that experimental implications as well as conceptual issues may hint to physics beyond this paradigm. Thus, we introduce in Sections \ref{sec:bounce1} and \ref{sec:bounce2} a class of very early universe models which are promising in addressing issues existing in inflationary cosmology. In particular, we present two representative paradigms. In Section \ref{sec:bounce1} we study in detail the new matter bounce scenario in which the universe starts with a matter-dominated contraction phase and evolves into an ekpyrotic phase and finally bounces into a thermal expansion. We review the predictions of this model on primordial perturbations and confront them with the latest cosmological observations. Afterwards, we point out that the simplest model suffers from the issue of the non-red tilt spectrum, and around this point, we propose possible mechanisms of generating a reasonable red tilt for primordial curvature perturbations in this model. The second is the matter-bounce inflation scenario as will be investigated in Section \ref{sec:bounce2}. This paradigm can be viewed as an extension of the regular inflationary cosmology with a matter contraction before inflation. In this section we study its phenomenological implications on the present cosmological observations. Particularly, we show explicitly that the models within this scenario generally have significant advantages in explaining simultaneously the Planck and BICEP2 observations better than the standard $\Lambda$CDM model, and are sensitive to be verified by the future CMB polarization measurements. Then, we briefly introduce a class of model constructions that can realize this scenario from effective field approach. Eventually, we conclude our work by addressing some unsettled issues in Section \ref{sec:conclusion}. Throughout the article we take the normalization of natural units and define the reduced Planck mass by $M_p = 1/\sqrt{8\pi G}$.

\section{General discussion on inflationary cosmology}\label{sec:inflation}

We begin with a brief discussion on general predictions of inflationary perturbations within the framework of General Relativity. Inflationary cosmology has become the most prevailing paradigm for describing the very early universe based on two reasons. Firstly, it can resolve some conceptual issues of the hot big bang cosmology. Secondly, it was the first model to provide a causal mechanism for generating primordial curvature perturbations which can explain the formation of the LSS \cite{Mukhanov:1981xt, Press:1980zz} and to interpret the existence of primordial gravitational waves in the CMB \cite{Grishchuk:1974ny, Starobinsky:1979ty}. More specifically, the perturbation theory developed in this paradigm predicted nearly scale-invariant power spectra of primordial perturbations both of scalar and tensor type which are highly Gaussian and adiabatic with a slightly red tilt. Most of these predictions have been accurately confirmed in a number of cosmological observations in the past decades of years \cite{Smoot:1992td, Bennett:2003bz, Abazajian:2003jy, Ade:2013zuv, Ade:2014xna}.

The inflationary perturbation theory was comprehensively reviewed in \cite{Mukhanov:1990me}. In this picture, both the primordial density perturbations and gravitational waves were originated from quantum fluctuations of the spacetime metric in a nearly exponentially expanding universe at high energy scales. During this exponential expansion, the physical wavelengthes of metric fluctuations can be stretched out of the Hubble radius and then form classical perturbations as observed by the CMB surveys. Such a convenient causal mechanism can spectacularly determine the power spectra of primordial perturbations of scalar and tensor types by a series of simple relations.

Specifically, we would like to start by considering a single field inflation model with a K-essence Lagrangian \cite{ArmendarizPicon:1999rj, Garriga:1999vw}. The corresponding power spectrum of primordial curvature perturbations is associated with four inflationary parameters, the Hubble rate $H$, the spectral index $n_s$, the slow roll parameter $\epsilon$, and the sound speed $c_s$ ($c_s=1$ when the Lagrangian is canonical), via the following form of parametrization
\begin{eqnarray}\label{PS}
 P_S = A_S \left( \frac{k}{k_{\rm pivot}} \right)^{n_S-1} ~,
\end{eqnarray}
with
\begin{eqnarray}\label{AS}
 A_S = \frac{H_I^2}{8\pi^2\epsilon c_s M_p^2}~,
\end{eqnarray}
where $k_{\rm pivot}$ is the pivot scale and $A_s$ is the spectrum amplitude of curvature perturbations. The subscript $_I$ denotes that the value of the Hubble parameter is taken during the inflationary phase.

The slow roll parameter $\epsilon$ is defined by
\begin{eqnarray}\label{epsilon}
 \epsilon \equiv -\frac{\dot{H}}{H^2}~,
\end{eqnarray}
and hence is determined only by the background dynamics of the inflaton field. In order to realize a sufficiently long inflationary phase, $\epsilon$ is required to be much less than unity. The sound speed parameter $c_s$ characterizes the propagating rate of the inflaton field fluctuations. Theoretically, its value is constrained between $0$ and $1$ so that the model is free from the gradient instability and super-luminal propagation. Moreover, the recent no-detection of primordial non-gaussianity by the Planck data \cite{Ade:2013ydc} strongly indicates that $c_s$ cannot be very small. The spectral index $n_S$ can be derived straightforwardly from its definition through
\begin{eqnarray}\label{nS}
 n_S-1 \equiv \frac{{\rm d}\ln P_S}{{\rm d} \ln k} = -4\epsilon+2\eta -s ~,
\end{eqnarray}
with two other slow roll parameters
\begin{eqnarray}\label{eta&s}
 \eta \equiv \epsilon - \frac{\dot{\epsilon}}{2H\epsilon}~, ~~ s\equiv \frac{\dot c_s}{Hc_s}~,
\end{eqnarray}
being introduced.
According to the current CMB observation, the spectral index $n_s$ takes a value which is slightly less than unity and hence, the power spectrum of primordial curvature perturbations is red tilted.

For primordial tensor fluctuations, the associated relations are even simpler if the gravity theory is General Relativity. The corresponding power spectrum takes the form of
\begin{eqnarray}\label{PTi}
 P_T = A_T \left( \frac{k}{k_{\rm pivot}} \right)^{n_T} ~,
\end{eqnarray}
with
\begin{eqnarray}\label{AT}
 A_T = \frac{2H_I^2}{\pi^2 M_p^2}~,
\end{eqnarray}
where the coefficient $A_T$ is the amplitude of tensor spectrum at the pivot scale. It is easy to see that the amplitude of primordial tensor fluctuations only depends on the inflationary Hubble parameter and the corresponding spectral index $n_T$, the latter of which, by definition, is given by
\begin{eqnarray}\label{nTi}
 n_T \equiv \frac{{\rm d}\ln P_T}{{\rm d} \ln k} = -2\epsilon~.
\end{eqnarray}
The expressions \eqref{PTi} and \eqref{nTi} are generic to any inflation models which minimally couple to General Relativity. Furthermore, it is convenient to define the tensor-to-scalar ratio as follows,
\begin{eqnarray}\label{ttsratio}
 r \equiv \frac{A_T}{A_S} = 16 c_s \epsilon ~,
\end{eqnarray}
which also characterizes the magnitude of primordial tensor fluctuations.

Based on these relations, one can make use of observational data to constrain the parameters introduced above and then further discriminate inflation models. Namely, according to the Planck observation of the CMB, the amplitude of the power spectrum of primordial curvature perturbations is constrained to be \cite{Ade:2013zuv}
\begin{eqnarray}
 \ln(10^{10}A_S) = 3.089^{+0.024}_{-0.027} ~(1\sigma ~{\rm CL})~,
\end{eqnarray}
at the pivot scale $k=0.002\,{\rm Mpc}^{-1}$. Moreover, the recently released BICEP2 result requires that \cite{Ade:2014xna}
\begin{eqnarray}
 r = 0.20^{+0.07}_{-0.05} ~(1\sigma ~{\rm CL})~.
\end{eqnarray}

However, from the historical perspective, it was known even before inflation that a roughly scale-invariant and almost adiabatic spectrum of cosmological perturbations could be a reasonable interpretation for the distribution of galaxies historically \cite{Peebles:1970ag, Sunyaev:1970eu}, which is now known as the famous ``Harrison-Zel'dovich" (HZ) power spectrum \cite{Harrison:1969fb, Zeldovich:1972zz}. Moreover, despite the phenomenological accomplishment, there exist some critical conceptual issues of inflationary cosmology that deserve to be treated seriously nowadays.

For one thing, inflationary cosmology does not resolve the problem of the initial singularity inherited from the hot big bang cosmology \cite{Borde:1993xh}. On the other hand, it is well known that the Planck mass suppressed corrections to the inflaton potential generally lead to the masses of the order of the Hubble scale and then spoil the slow roll conditions rendering a sustained inflationary stage impossible \cite{Copeland:1994vg}. This issue would be even worse if the field variation of the inflaton is super-Planckian as indicated by the BICEP2 \cite{Lyth:1996im}. Moreover, if we trace backward along the cosmological perturbations observed today, their length scales could go beyond the Planck length at the onset of inflation \cite{Brandenberger:1999sw, Martin:2000xs}. Additionally, in order to study quantum field theory during inflation, it is inevitably necessary to systematically study the nonlinear corrections of field fluctuations that are on one side not ultraviolet (UV) complete, and on the other side yield observably large infrared effects that were not detected in experiments \cite{Tsamis:1992sx, Mukhanov:1996ak}.

Keeping the above theoretical remarks and the previous phenomenological motivations in mind, it is worth looking for plausible alternative paradigm that might not only be as successful as inflation in phenomenologically explaining the CMB and LSS of our universe, but also can resolve or at least circumvent some conceptual issues mentioned above. In the following sections, we will present two interesting paradigms.

\section{Toward the big bang in bouncing cosmology: the paradigm of matter-ekpyrotic bounce}\label{sec:bounce1}

Non-singular bouncing cosmologies can resolve the initial singularity problem of the inflationary $\Lambda$CDM model and hence have attracted a lot of attention in the literature. They appear in many theoretical settings where either the gravitational sector is modified as in Ho\u{r}ava gravity \cite{Calcagni:2009ar, Kiritsis:2009sh, Brandenberger:2009yt}, non-relativistic gravitational action\cite{Cai:2009in}, torsion gravity \cite{Cai:2011tc, Poplawski:2011jz}, Lagrange-multiplier gravity \cite{Cai:2010zma, Cai:2011bs}, nonlinear massive gravity \cite{Cai:2012ag} and non-local gravity \cite{Biswas:2005qr}, or by making use of matter fields with the NEC violation such as in the quintom bounce \cite{Cai:2007qw, Cai:2007zv}, the Lee-Wick bounce \cite{Cai:2008qw, Bhattacharya:2013ut}, the ghost condensate bounce \cite{Buchbinder:2007ad, Creminelli:2007aq, Lin:2010pf}, and the S-brane bounce \cite{Brandenberger:2013zea} models. A non-singular bounce may also be achieved in a universe with non-flat spatial geometry (see e.g. \cite{Martin:2003sf, Solomons:2001ef}). We refer to \cite{Novello:2008ra, Lehners:2008vx} for recent reviews of various bouncing cosmologies. However, the BKL instability appears in contracting cosmologies as the effective energy density contributed by the back-reaction of anisotropies increases faster than the dust and radiation densities, unless one finely tunes the initial conditions to be nearly perfectly isotropic in order to ensure that anisotropies never dominate. However, this issue is avoided in the ekpyrotic scenario where a scalar field with a steep and negative-valued potential always dominates over anisotropies in a contracting universe \cite{Erickson:2003zm} and so it is justified to neglect anisotropies in the presence of an ekpyrotic scalar field.

Recently, it has been shown how it is possible to combine an era of ekpyrotic contraction with a non-singular bounce by introducing a scalar field with a Horndeski-type non-standard kinetic term and a negative exponential potential \cite{Cai:2012va}. Furthermore, one may include a regular dust field and assume the universe began in a state of matter-dominated contraction thus combining the matter bounce with the ekpyrotic scenario. It can be explicitly checked that anisotropies remain small throughout the entire cosmological evolution of the matter-ekpyrotic bounce model, including at the bounce point \cite{Cai:2013vm}, and therefore this model successfully avoids the BKL instability that arises for a large family of non-singular bounce models, as pointed out in \cite{Xue:2010ux, Xue:2011nw}. Among many possible implementations of the matter bounce, a concrete realization of the matter-ekpyrotic bounce was constructed in \cite{Cai:2013kja} that involves two matter fields with one being the scalar field that causes the bounce and the other representing the matter field that is dominant at the beginning of the contracting phase. Note that this effective field theory model of a non-singular bounce can be developed into a supersymmetric version \cite{Koehn:2013upa}, or embedded into loop quantum cosmology (LQC) \cite{Cai:2014zga} (also see the LQC realizations of matter bounce in \cite{WilsonEwing:2012pu}, ekpyrotic bounce in \cite{Wilson-Ewing:2013bla}, and matter-bounce inflation in \cite{Amoros:2014tha} separately). In the present section, we shall study the matter-ekpyrotic bounce within the context of effective field approach and its cosmological implications to observations.

\subsection{The model building with two fields}\label{subsec:nmb}

We briefly review how it is possible to obtain a non-singular bounce and avoid the BKL instability in the new matter bounce cosmology. We phenomenologically consider a background model involving two scalar fields with a Lagrangian of the following type \cite{Cai:2013kja}:
\begin{align} \label{L_KGB}
 {\cal L} = K(\phi, X) + G(\phi, X)\Box\phi + {\cal L}_\chi~,
\end{align}
where $\phi$ is the bounce field and $\chi$ is a regular matter field. Specifically we choose the operators $K$ and $G$ of the bounce field to be
\begin{align} \label{KG}
 K &= [1-g(\phi)] X +\frac{\beta X^2}{M_p^4} -V(\phi)~, \\
 G &= \frac{\gamma X}{M_p^3} ~,
\end{align}
with $X$ being defined as the regular kinetic term $X \equiv g^{\mu\nu} (\partial_\mu\phi) (\partial_\nu\phi) /2$, while $\beta$ and $\gamma$ are coupling constants and $\Box \equiv g^{\mu\nu}\nabla_\mu\nabla_\nu$ is the standard d'Alembertian operator.

Note that in this model the $K$ operator involves the term $\beta X^2$ which can stabilize the kinetic energy of the scalar field at high energy scales when $\beta$ is positive-definite. A bouncing phase can be triggered by allowing $g$ to become larger than unity for a short while which leads to the emergence of a ghost condensate. The value of $g$ is required to be small far away from the bouncing phase so that the kinetic term in the Lagrangian is well approximated by the standard kinetic term before and after the bounce. Additionally, the potential $V(\phi)$ governs the dynamics of $\phi$ away from the bounce as well as determines the energy scale the bounce occurs at. In order to dilute the unwanted anisotropies to avoid the BKL instability, one takes the potential to have the ekpyrotic form of a negative exponential (at least for $\phi \ll -M_p$). To be specific, following \cite{Cai:2012va, Cai:2013kja} we take the function $g(\phi)$ and the potential $V(\phi)$ to be
\begin{align}
 g(\phi) &= \frac{2g_0}{e^{-\sqrt{\frac{2}{p_g}}\frac{\phi}{M_p}} +e^{b_g\sqrt{\frac{2}{p_g}}\frac{\phi}{M_p}}}~, \\
 \label{gV}
 V(\phi) &= -\frac{2V_0}{e^{-\sqrt{\frac{2}{q}}\frac{\phi}{M_p}} +e^{b_V\sqrt{\frac{2}{q}}\frac{\phi}{M_p}}}~,
\end{align}
where $p_g, q, b_g, b_V$ and $g_0\equiv g(0)$ are dimensionless positive constants and $V_0$ is also positive with dimensions of $({\rm mass})^4$. We choose $g_0$ to be slightly larger than unity so that the scalar can form a ghost condensate state when $\phi$ is near zero. The critical value of $g$ that signals the onset of the non-singular bouncing phase is therefore $g(\phi_*)=1$. By solving this equation, one finds the approximate values of $\phi_*$ where the non-singular bouncing phase begins and ends, respectively $\phi_{*-} \simeq -M_p\ln (2g_0 / p_g) $ and $\phi_{*+} \simeq M_p\ln (2g_0 /b_g p_g)$. Moreover, in order to obtain ekpyrotic contraction, we take $q < 1$.

\subsubsection{The phase of matter contraction}
Since the second scalar $\chi$ is viewed as a regular matter field, we may simply take the Lagrangian to be that of a free canonically normalized massive scalar field:
\begin{eqnarray}
 {\cal L}_\psi = \frac{1}{2}\partial_\mu\chi\partial^\mu\chi -\frac{1}{2}m^2\chi^2~.
\end{eqnarray}
Since $\chi$ oscillates around its vacuum state $\chi=0$, the time-averaged background equation-of-state parameter is $w=0$. At the beginning of the cosmic evolution, if the universe is dominated by $\chi$, then the scale factor evolves as
\begin{eqnarray}
 a(t) \simeq a_{E} \left( \frac{t-\tilde{t}_{E}}{t_{E} -\tilde{t}_{E}} \right)^{2/3},
\end{eqnarray}
where $t_{E}$ denotes the final moment of matter contraction and the beginning of the Ekpyrotic phase, and $a_{E}$ is the value of the scale factor at the time $t_{E}$. In the above, $\tilde{t}_{E}$ is an integration constant which is introduced to match the Hubble parameter continuously at the time $t_{E}$, i.e., $\tilde{t}_{E} \simeq t_{E}-\frac{2}{3H_{E}}$. In this phase the time-averaged Hubble parameter can be approximated by
\begin{eqnarray}\label{Hubble_c}
 \langle H(t) \rangle = \frac{2}{3(t-\tilde{t}_{E})}~.
\end{eqnarray}
where the angular brackets stand for averaging over time.

\subsubsection{The phase of Ekpyrotic contraction}
We assume a homogeneous scalar field $\phi$ which is initially placed in the region $\phi \ll -M_p$ in the phase of matter contraction. In this case, the Lagrangian for $\phi$ approaches the conventional canonical form. Once $\phi$ begins to dominate the universe, the equation-of-state parameter then approaches an Ekpyrotic one
\begin{eqnarray}\label{eos_ekpy}
 w \simeq -1+\frac{2}{3q} ~.
\end{eqnarray}
During the phase of Ekpyrotic contraction, the scale factor evolves as
\begin{eqnarray}
 a(t) \simeq a_{B-} \left(\frac{t-\tilde{t}_{B-}}{t_{B-}
 -\tilde{t}_{B-}}\right)^q ~,
\end{eqnarray}
where $\tilde{t}_{B-} = t_{B-}-\frac{q}{H_{B-}} $ is chosen such that the Hubble parameter at the end of the phase of Ekpyrotic contraction matches with the one at the beginning of the bounce phase, and $a_{B-}$ is the value of scale factor at the time $t_{B-}$. Therefore, the Hubble parameter is given by
\begin{eqnarray}\label{Hubble_E}
 H(t) \simeq \frac{q}{t-\tilde{t}_{B-}} ~.
\end{eqnarray}
Additionally, we require the scale factor to evolve smoothly and continuously at the time $t_{E}$. This leads to the relation
\begin{eqnarray}\label{a_E}
 a_{E} \simeq a_{B-} \left(\frac{H_{B-}}{H_{E}}\right)^q ~.
\end{eqnarray}

\subsubsection{The nonsingular bouncing phase}
In this model the scalar field evolves monotonically from $\phi \ll -M_p$ to $\phi \gg M_p$. When $\phi$ evolves into the regime between $\phi_{*-}$ and $\phi_{*+}$, the coefficient $g(\phi)$ becomes larger than unity and thus the universe enters a ghost condensate state. As shown in Ref. \cite{Cai:2012va}, during this phase, the Hubble parameter evolves approximately as
\begin{eqnarray}\label{Hubble_bounce}
 H(t) \simeq \Upsilon t ~,
\end{eqnarray}
as well as the background scalar
\begin{eqnarray}\label{dotphi_bouncing}
 \dot\phi(t) \simeq \dot\phi_{B} e^{-t^2/T^2} ~,
\end{eqnarray}
where the coefficient $\Upsilon$ is set by the detailed microphysics of the bounce. The coefficient $T$ can be determined by matching the detailed evolution of the scalar field at the beginning or the end of the bounce phase. Thus, during the bounce the scale factor evolves as
\begin{eqnarray}
 a(t) \simeq a_{B} e^{\Upsilon t^2/2} ~,
\end{eqnarray}
which is exactly a nonsingular bouncing solution.

\subsubsection{The phase of fast-roll expansion}
After the bounce, the universe enters the expanding phase, where the universe is still dominated by the scalar field $\phi$. During this stage, the motion of $\phi$ is dominated by its kinetic term while the potential is negligible. Thus, the background equation-of-state parameter is $w \simeq 1$. This corresponds to a period of fast-roll expansion, where the scale factor evolves as
\begin{eqnarray}
 a(t) \simeq a_{B+} \left(\frac{t-\tilde{t}_{B+}}
 {t_{B+}-\tilde{t}_{B+}}\right)^{1/3} ~,
\end{eqnarray}
where $t_{B+}$ represents the end of the bounce phase and the beginning of the fast-roll period, and $a_{B+}$ is the value of the scale factor at that moment. Then one can write down the Hubble parameter in the fast-roll phase
\begin{eqnarray}
 H(t) \simeq \frac{1}{3(t-\tilde{t}_{B+})} ~,
\end{eqnarray}
and the continuity of the Hubble parameter at $t_{B+}$ yields $\tilde{t}_{B+} = t_{B+} -\frac{1}{3H_{B+}}$. By virtue of this phase, it is then natural to connect to the thermal expanding history of the big bang cosmology.

\subsubsection{Numerics}
\begin{figure}
\includegraphics[scale=0.3]{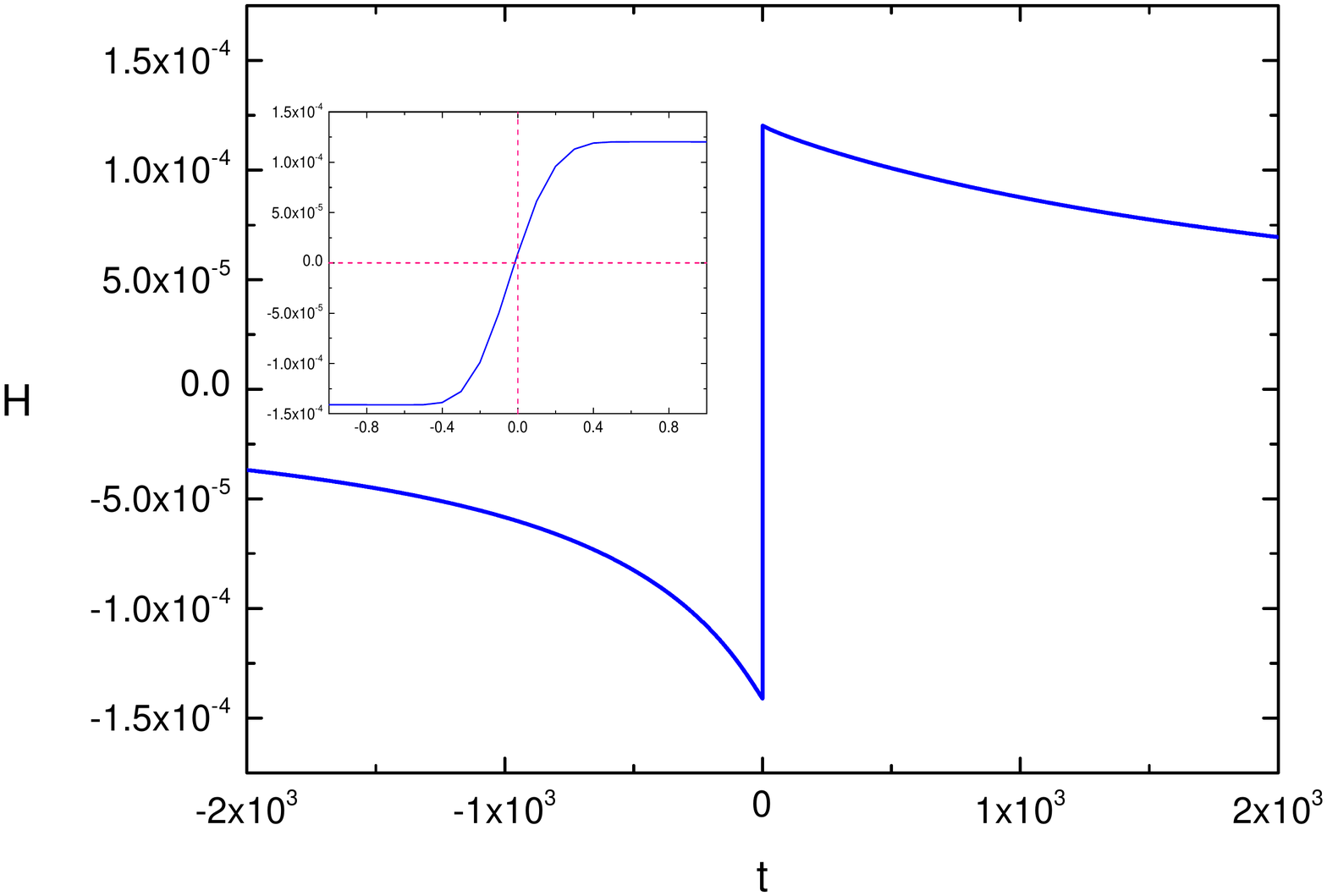}
\includegraphics[scale=0.3]{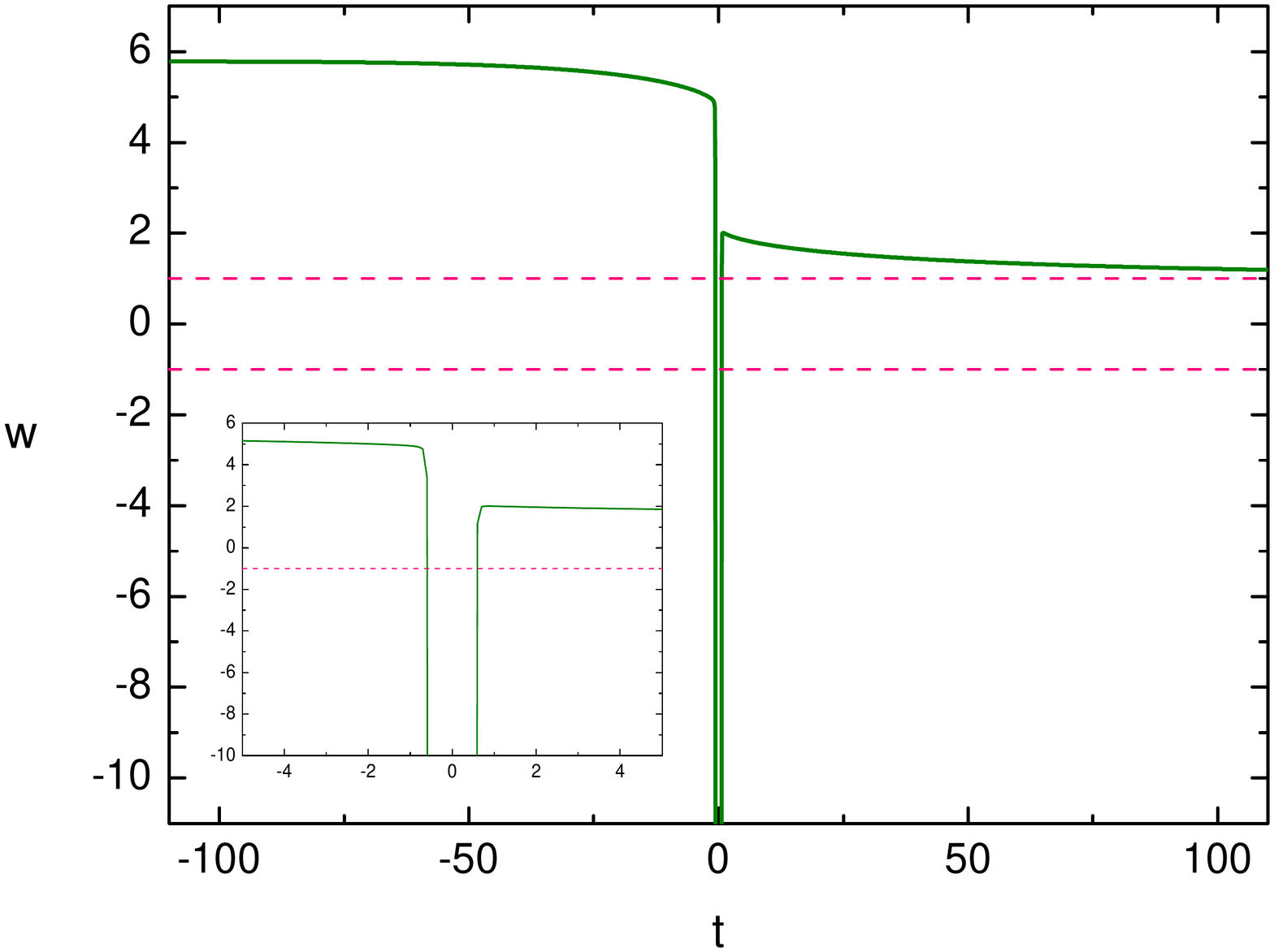}
\caption{Numerical plot of the Hubble parameter $H$ and the equation-of-state parameter $w$ as a function of cosmic time in the new matter bounce cosmology. The insert shows the detailed evolution around the bounce point. Planck units are used. From \cite{Cai:2012va}.}
\label{Fig:EoS}
\end{figure}

In Fig. \ref{Fig:EoS}, we numerically plot the evolution of the Hubble parameter and the equation-of-state in the new matter bounce paradigm, respectively. Also shown are zoomed-in views of the evolution around the bounce point. One can see from the upper panel of Fig. \ref{Fig:EoS} that the Hubble parameter $H$ evolves smoothly through the bounce point with a dependence on cosmic time which is close to linear. In the lower panel of the figure, one can find that $w$ in the contracting phase can be larger than unity which corresponds to an ekpyrotic phase. If one traces backward to an earlier time, then there would be a matter-like contracting phase which is not presented in the plot. In addition, after the bounce, the equation-of-state parameter quickly approaches the value $w = 1$ which corresponds to the kinetic-driven phase of expansion. During this period the contribution of the bounce field $\phi$ will be diluted quickly relative to the contributions of regular matter and radiation. Thus, the universe in this model can connect smoothly to the thermal expansion of the hot big bang cosmology.

\subsection{Observational constraints}\label{subsec:nmb_constraint}

It is known that, both in inflation or matter bounce cosmology, the amplitudes of scalar and tensor fluctuations are originally comparable. For inflation models, the scalar perturbations can achieve an enhancement due to the slow roll parameter. For matter bounce cosmologies, the tensor-to-scalar ratio is in general predicted to be of order unity since the slow roll condition is no longer required in this scenario. Thus, at first glance, this issue is disturbing for a large class of matter bounce cosmologies \cite{Cai:2008qw}. However, it also implies a theoretical upper bound for the tensor-to-scalar ratio in the matter bounce paradigm. If there is any mechanism that can amplify scalar perturbations, then one can always easily suppress this ratio to fit to observations. For example, in some explicit models this value can be suppressed due to the nontrivial physics of the bouncing phase, namely, the matter bounce curvaton \cite{Cai:2011zx} and the new matter bounce cosmology \cite{Cai:2013kja}.

In the new matter bounce model introduced in the previous subsection, we have introduced two scalar fields with $\chi$ driving a phase of matter contraction and $\phi$ being responsible for the ekpyrotic contraction and the bounce. Correspondingly, there exist primordial curvature perturbations and entropy fluctuations during the matter-dominated contracting phase. The fluctuations seeded by $\phi$, which are the entropy modes originally, can be converted into curvature perturbations when the universe enters the ekpyrotic phase. Therefore, the power spectrum of cosmological perturbations in this model is highly adiabatic after the bounce. In this model, when the universe evolves into the bouncing phase, the kinetic term of the scalar field that triggers the bounce could vary rapidly. This process also effectively leads to a tachyonic-like mass for curvature perturbations, and hence, the amplitude can be amplified. Correspondingly, the tensor-to-scalar ratio is suppressed when primordial perturbations pass through the bouncing phase in the new matter bounce cosmology.

The observational constraint of this scenario was performed in detail in \cite{Cai:2014xxa}. To be specific, the power spectrum for primordial gravitational waves is expressed as
\begin{eqnarray}
 P_T \simeq \frac{{\cal F}_{\psi}^2\gamma_\psi^2 H_E^2}{16\pi^2(2q-3)^2M_p^2}~,
 \label{PT}
\end{eqnarray}
with
\begin{align}
 \gamma_\psi &\simeq \frac{1}{2(1-3q)}~, \nonumber\\
 {\cal F}_{\psi} &\simeq \exp \left[ 2\sqrt{\Upsilon}t_{B+} +\frac{2}{3}\Upsilon^{3/2}t_{B+}^3 \right]~,
\end{align}
at leading order. In the above expression, we have applied the assumption that the values of the scale factor before and after the bounce are comparable. We denote the time at the beginning and the end of the bouncing phase by $t_{B-}$ and $t_{B+}$, respectively. Up to leading order, the power spectrum of curvature fluctuations is given by
\begin{equation}
 P_\zeta\simeq\frac{\mathcal{F}_{\zeta}^2H_E^2a_E^2}{8\pi^2M_p^4}\gamma_{\zeta}^2m^2\left|U_{\zeta}\right|^2~,
 \label{Pz}
\end{equation}
with $\gamma_\zeta\simeq\gamma_\psi$ and
\begin{align}
  U_\zeta &= -(25+49q)i\frac{H_E}{24m}-\frac{27q}{24}~,\nonumber \\
  {\cal F}_{\zeta} &\simeq e^{2\sqrt{2+\Upsilon T^2}\left(\frac{t_{B+}}{T}\right) +\frac{2(2+3\Upsilon T^2+\Upsilon^2T^4)}{3\sqrt{2+\Upsilon T^2}} \left(\frac{t_{B+}^3}{T^3}\right)}~.
\end{align}
Combining Eqs.\ (\ref{PT}) and (\ref{Pz}), one can write down the tensor-to-scalar ratio in this model as follows,
\begin{equation}
 r \equiv\frac{P_T}{P_\zeta}
 \simeq \frac{{\cal F}_{\psi}^2M_p^2}{2(2q-3)^2 \mathcal{F}_{\zeta}^2a_E^2m^2 \left| U_{\zeta}^{(k)} \right|^2}~.
\end{equation}

In Ref. \cite{Cai:2014xxa}, the observational constraints on the parameter space of the new matter bounce scenario have been investigated numerically. It is interesting to notice that, in this model there are only three main parameters that are most sensitive to observations, which are, the slope parameter $\Upsilon$, the Hubble rate at the beginning of the ekpyrotic phase $H_E$, and the dimensionless duration parameter $t_{B+}/T$ of the bouncing phase.

\begin{figure}
\includegraphics[scale=0.6]{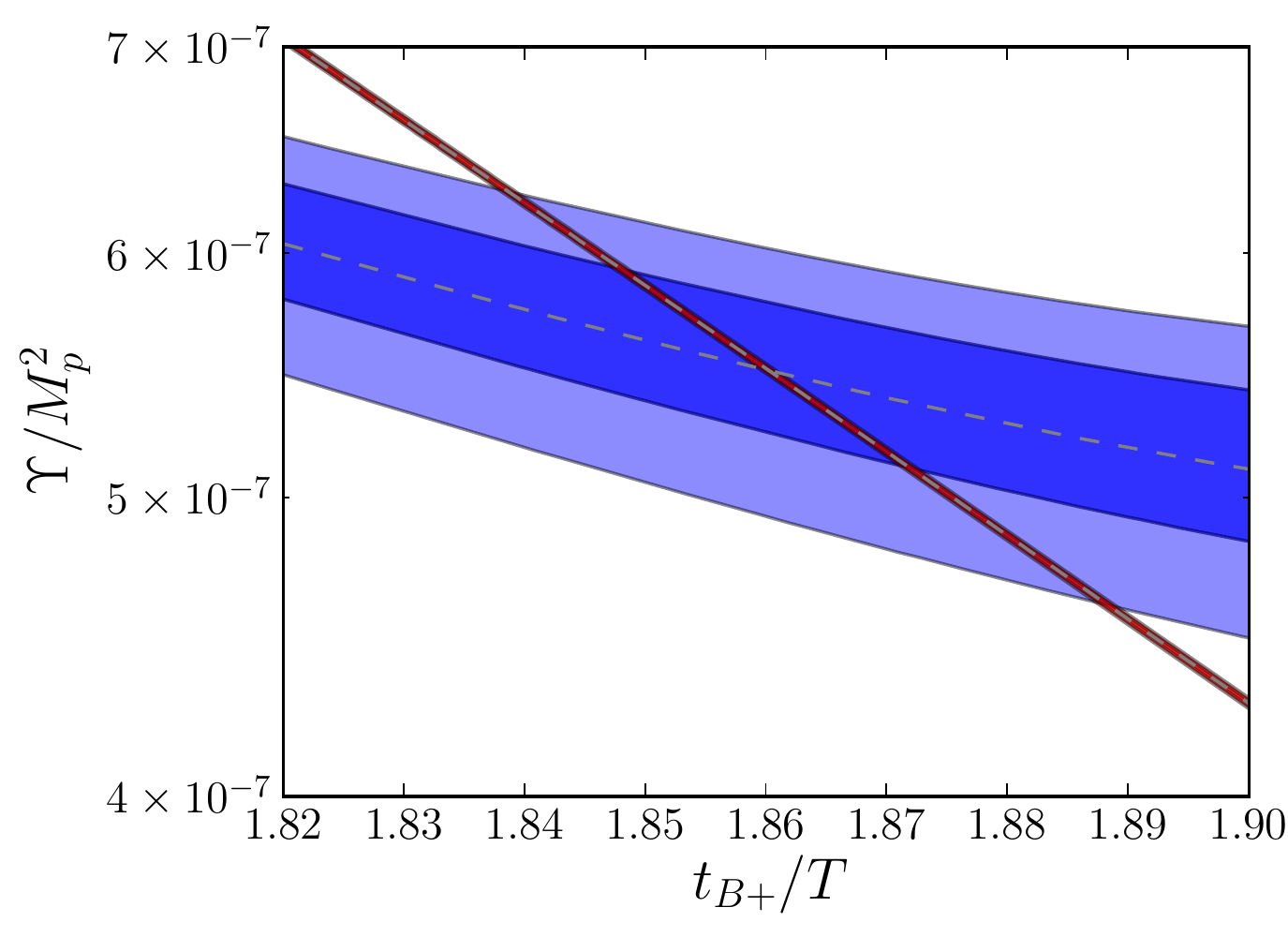}
\caption{Constraints on the dimensionless duration parameter $t_{B+}/T$ and the slope parameter $\Upsilon$ of the new matter bounce cosmology from the combined Planck and BICEP2 data. The value of the Hubble parameter at the beginning of the ekpyrotic phase is fixed to be $H_E/M_p=10^{-7}$. The blue bands show the $1\sigma$ and $2\sigma$ confidence intervals of the tensor-to-scalar ratio and the red bands show the confidence intervals of the amplitude of the power spectrum of curvature perturbations. From \cite{Cai:2014xxa}. }
\label{Fig:nmb2}
\end{figure}

The correlation between $\Upsilon$ and $t_{B+}/T$ is shown in Fig. \ref{Fig:nmb2}. From the figure, one can find that $\Upsilon$ and $t_{B+}/T$ are slightly negatively correlated. This implies that one expects either a slow bounce with a long duration or a fast bounce with a short duration. In addition, the constraint on the dimensionless duration parameter $t_{B+}/T$ is very tight with a value slightly less than $2$. Thus, it is important to examine whether the model predictions accommodate with observations after choosing a fixed value of $t_{B+}/T$.

\begin{figure}
\includegraphics[scale=0.6]{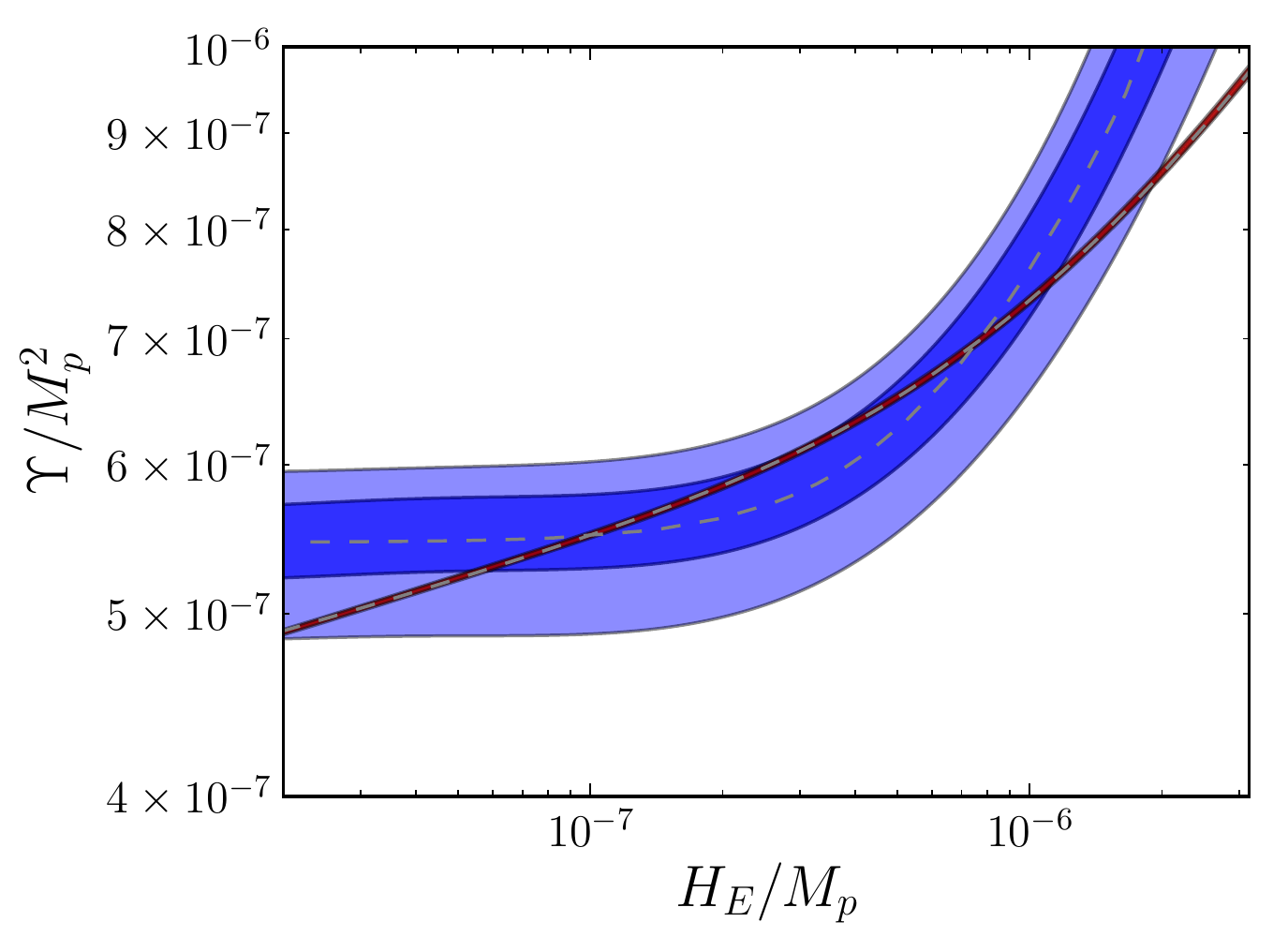}
\caption{Constraints on the Hubble parameter $H_E$ and the slope parameter $\Upsilon$ of the new matter bounce cosmology from the combined Planck and BICEP2 data. The dimensionless bounce time duration is fixed to be $t_{B+}/T=1.86$. As in Fig. \ref{Fig:nmb2}, the blue and red bands show the confidence intervals of $r$ and $P_\zeta$, respectively. From \cite{Cai:2014xxa}. }
\label{Fig:nmb5}
\end{figure}

Afterwards, we analyze the correlation between $\Upsilon$ and $H_E$ by setting $t_{B+}/T=1.86$. The allowed parameter space is depicted by the intersection of the blue and red bands shown in Fig. \ref{Fig:nmb5}. From the numerical result, we find that the Hubble scale $H_E$ and the slope parameter $\Upsilon$ introduced in the new matter bounce cosmology are constrained to be in the following ranges
\begin{align}
 1.9 \times 10^{-8} &\lesssim H_E/M_p \lesssim 1.9 \times 10^{-6} ~, \\
 4.9 \times 10^{-7} &\lesssim \Upsilon/M_p^2 \lesssim 8.5 \times 10^{-7} ~,
\end{align}
respectively. It is easy to see that, for the new matter bounce cosmology, if we assume that the bounce occurs at the GUT scale, then the duration of the bouncing phase is roughly $\mathcal{O}(10^4)$ Planck times. Also, according to the constraint the amplitude of the Hubble scale $H_E$ has to be lower than the GUT scale. This allows for a sufficiently long ekpyrotic contracting phase that can suppress the unwanted primordial anisotropies as addressed in \cite{Cai:2013vm}.

From the analysis of the new matter bounce cosmology, we can conclude that, a nonsingular bouncing cosmology has to experience the bouncing phase smoothly enough to be consistent with the latest CMB observations. For example, the Hubble parameter cannot grow very fast during the bounce since the constraints that we find favor a small value of $\Upsilon$. In the model we considered, the observed amplitude of the spectra of CMB fluctuations is determined by the Hubble parameter at the moment when primordial perturbations were frozen at super-Hubble scales, i.e., $H_E$ at the onset of the ekpyrotic phase, which is required to be much lower than the GUT scale in order to agree with observations. These interesting results encourage further studies of bouncing cosmologies following the effective field approach.

\subsection{Mechanisms for generating a red tilt for primordial perturbations}\label{subsec:nmb_redtilt}

In the paradigm of the new matter bounce cosmology, a free scalar field can yield a scale-invariant power spectrum. For example, in the model considered above, the bounce field $\phi$ gives rise to a scale-invariant power spectrum for entropy perturbations, which during the ekpyrotic phase are able to be converted into adiabatic modes. However, it is interesting to note that this spectrum is too close to the HZ spectrum rather than a red tilt spectrum as observed by WMAP and Planck. Therefore, it is important to look for suitable mechanisms capable of generating a red tilt for primordial perturbations in matter bounce cosmologies.

Before going to the detailed mechanisms, we would like to briefly explain how the perturbations form a scale-invariant power spectrum in the very early universe. A generic analysis on the conditions for generating scale-invariant primordial spectra in various cosmological evolutions was performed in \cite{Cai:2009hc}. Based on that work, we study in this subsection the corresponding cosmological implications.

To be specific, we investigate the possible ways of generating a power spectrum of primordial perturbations with a modified dispersion relation. To seed the fluctuations, a test scalar field $\varphi$ is introduced which does not affect the background evolution in primordial stages, but yields a new degree of freedom for scalar perturbations $\delta\varphi$. In Fourier space, the equation of motion for its fluctuation $\delta\varphi$ is given by
\begin{eqnarray}\label{eom_p}
 v_k''+(\nu^2-\frac{a''}{a})v_k=0~,
\end{eqnarray}
where the variable $v_k$ is the Mukhanov-Sasaki variable defined by $v_k\equiv a\delta\varphi_k$ \cite{Sasaki:1986hm, Mukhanov:1988jd}, and the prime denotes the derivative with respect to the comoving time $\eta\equiv\int {dt}/{a}$. For a constant equation-of-state $w$, there is
\begin{eqnarray}\label{mass}
 \frac{a''}{a}= \frac{p(2p-1)}{(1-p)^2} \frac{1}{\eta^2}~,
\end{eqnarray}
where $p=2/3(1+w)$ has been introduced. Moreover, $\nu(k)$ is introduced as a generalized frequency for the perturbation mode. Since the dispersion relation may be modified at high energy scales, and if this happens, one could phenomenologically consider a generalized dispersion relation
\begin{eqnarray}\label{mdf}
 \nu = k ~f(k_{ph})~,
\end{eqnarray}
with
\begin{eqnarray}
f(k_{ph}) =
  \left\{ \begin{array}{c}
          (\frac{k_{ph}}{M})^\alpha ~,~~ k_{ph}>M \\
          \\
          ~~ 1 ~~~~~~,~~ k_{ph} \leq M
\end{array} \right.
\end{eqnarray}
where $k_{ph}={k}/{a}$ is a physical wave number and $\alpha$ is a positive-definite parameter. The energy scale $M$ is assumed to be associated with new physics beyond which the standard dispersion relation is modified.

From Eq. (\ref{eom_p}), it is obvious that only two terms would affect the solution, which are $\nu^2$ and $a''/a$, respectively. Firstly, let us neglect the term $\frac{a''}{a}$ and focus on the asymptotic solution under the limit $|\nu\eta|\gg1$. This asymptotic solution oscillates like a sine function. This feature coincides with an adiabatic condition $|\nu'/\nu^2|\gg1$, which corresponds to the case in which the effective physical wavelength is deep inside the Hubble radius. Therefore, the modes can be regarded as adiabatic when they are staying in the sub-Hubble regime with $|\nu\eta|\gg1$, and we may impose a suitable initial condition by virtue of a Wentzel-Kramers-Brillouin (WKB) approximation
\begin{eqnarray}\label{inicond}
 v_k^{i}\simeq\frac{1}{\sqrt{2\nu(k,\eta)}}e^{-i\int^\eta\nu(k,\tilde\eta)d\tilde\eta}~.
\end{eqnarray}

Secondly, we look at the effective mass term $\frac{a''}{a}$ closely. Note that the generation of primordial fluctuations strongly depends on the background evolution. In suitable environments, the variable $|\nu\eta|$ could decrease along with the cosmic evolution. Once $|\nu\eta|\ll1$, the modes would exit the Hubble radius, and the equation of motion yields another asymptotic solution of which the leading term takes the form of
\begin{eqnarray}\label{sollead}
 v_k^{l} \sim \eta^{\frac{1}{2}} \bigg[ c(k)\eta^{-\frac{1}{2}|\frac{1-3p}{1-p(1+\alpha)}|} \bigg]~.
\end{eqnarray}

Afterwards, we match the above two asymptotic solutions (\ref{inicond}) and (\ref{sollead}) at the moment of Hubble crossing $|\nu\eta|\sim1$ and then determine the form of $v_k$ at super-Hubble scales as follows,
\begin{eqnarray}\label{solution_mdf}
 v_k(\eta) \simeq \frac{1}{\sqrt{2\nu}}
\bigg(\nu(k,\eta)\eta\bigg)^{\frac{1}{2}-\frac{1}{2}|\frac{1-3p}{1-p(1+\alpha)}|}~.
\end{eqnarray}
Applying the solution (\ref{solution_mdf}) directly, we obtain the primordial spectrum as follows:
\begin{eqnarray}\label{spectrum}
 P_{\delta\phi}&=&\frac{k^3}{2\pi^2}|\frac{v_k(\eta)}{a}|^2~\nonumber\\
  &\sim& \left\{ \begin{array}{c}
               k^{\frac{2-\alpha}{1-p-\alpha p}}~,~~(1-3p)[1-p(1+\alpha)]\geq0 \\
               \\
               k^{\frac{4+\alpha-6p(1+\alpha)}{1-p-\alpha
               p}}~,~(1-3p)[1-p(1+\alpha)]<0~.
\end{array} \right.
\end{eqnarray}

Finally, we obtain the power spectrum of primordial perturbations without considering any specific models. From the result (\ref{spectrum}), we find that there are four sufficient conditions for the spectrum to be scale-invariant (i.e., $P_{\delta\phi}\sim k^0$). They are:
\begin{itemize}
\item (i) $p\rightarrow+\infty$ in the first case, which corresponds to inflationary cosmology \cite{Guth:1980zm};
\item (ii) $p\rightarrow-\infty$ in the first case, which can be achieved in island cosmology\cite{Dutta:2005gt, Piao:2005na};
\item (iii) $\alpha=2$ in the first case, which can be obtained in the model of Ho\v{r}va-Lifshitz cosmology\cite{Calcagni:2009ar, Kiritsis:2009sh};
\item (iv) $p=\frac{4+\alpha}{6(1+\alpha)}$ in the second case.
\end{itemize}

Among these four conditions, the last one can be applied to bouncing cosmology, in particular, to the matter bounce scenario.
In the following we will study the spectrum of primordial perturbations in this case. Specifically, when the universe is in a matter-like contracting phase, one can fix the background equation-of-state parameter as $w=0$ and hence get $p={2}/{3}$. In this case, the contribution of $a''/a$ is also fixed. In this type of models, cosmological perturbations with a standard dispersion relation were analyzed in detail as shown in Refs. \cite{Cai:2008ed, Cai:2008qw}, and primordial non-Gaussianities were studied by Refs. \cite{Cai:2009fn, Cai:2009rd}. It turns out that the power spectrum of the IR modes is given by,
\begin{eqnarray}
 P_{\delta\phi}^{\rm IR} = \left( \frac{H_B}{4\pi} \right)^2~,
\end{eqnarray}
where $H_B$ is associated with the maximal Hubble scale around the bounce point.

\subsubsection{Varying $\nu^2$}

Now we deal with the UV modes of which the comoving wave numbers achieve a modification on their dispersion relation. For these modes, we can solve the perturbation equation and obtain the following solution:
\begin{eqnarray}
 v_k^{\rm UV}\simeq\frac{1}{\sqrt{2\nu}}(\nu\eta)^{-\frac{1+\alpha}{1-2\alpha}}~.
\end{eqnarray}
From this solution, one can read that the amplitude of the super-Hubble modes keeps growing until the bounce takes place, and thus we can calculate the power spectrum around the bounce point, which is given by
\begin{eqnarray}
 P_{\delta\phi}^{\rm UV} = \frac{1}{2^{\frac{2(2-\alpha)}{1-2\alpha}}\pi^2} H_B^{\frac{2(1+\alpha)}{1-2\alpha}} M^{\frac{3\alpha}{1-2\alpha}} \left( \frac{k}{a_B} \right)^{-\frac{9\alpha}{1-2\alpha}}~.
\end{eqnarray}
The corresponding spectral index can be derived as follows:
\begin{eqnarray}\label{n_phi_UV}
 n_\phi \equiv 1+\frac{d\ln P_{\delta\phi}}{d\ln k} = 1-\frac{9\alpha}{1-2\alpha}~,
\end{eqnarray}
which is red-tilted in the UV regime.
\begin{figure}[htbp]
\includegraphics[scale=0.8]{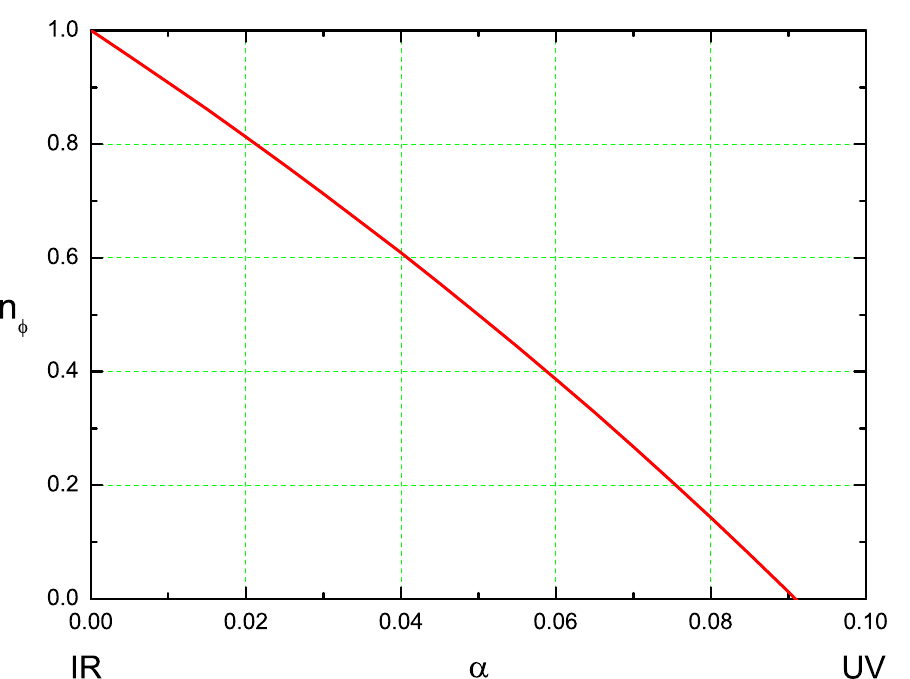}
\caption{The spectral index of the primordial perturbations with a modified dispersion relation as a function of the parameter $\alpha$ in the UV regime. From \cite{Cai:2009hc}.} \label{fig:nsnualpha}
\end{figure}
In Fig. \ref{fig:nsnualpha} we numerically plot the spectral index of the primordial perturbations described by the above mechanism. One may notice that, in order to fit the cosmological observations, e.g. the Planck data \cite{Ade:2013zuv}, the model parameter $\alpha$ cannot deviate from zero too much, which provides a strong constraint on models of modified gravity. However, this result successfully illustrates that a slightly red spectrum may be obtained by making a very small modification to the dispersion relation in a matter bounce model.

\subsubsection{Varying $a''/a$}

After having analyzed the possible effect brought by the effective frequency, now we turn to the study of how the effective mass term $a''/a$ affects the spectral index of the scalar spectrum. For convenience, we fix the frequency term by taking the regular dispersion relation $\nu = k$, and hence, $\alpha = 0$. In this case, we solve the perturbation equation and obtain the super-Hubble solution:
\begin{eqnarray}
 v_k^{\rm IR}\simeq\frac{1}{\sqrt{2k}}(k\eta)^{\frac{1-2p}{1-p}}~.
\end{eqnarray}
Substituting this solution into the expression of the scalar spectrum \eqref{spectrum}, one can derive
\begin{eqnarray}
 P_{\delta\phi}^{\rm IR} = \frac{1}{4\pi^2} (\frac{p}{1-p})^{\frac{2-4p}{1-p}} H_B^{\frac{4p-2}{1-p}} \left( \frac{k}{a_B} \right)^{\frac{4-6p}{1-p}} ~,
\end{eqnarray}
around the bounce point. It is easy to write down the spectral index as
\begin{eqnarray}
 n_\phi \equiv 1+\frac{d\ln P_{\delta\phi}}{d\ln k} = 1+\frac{4-6p}{1-p}~.
\end{eqnarray}
correspondingly. From this result, we can easily find that a red tilt can also be obtained if we choose $p$ to be slightly larger than $2/3$.
\begin{figure}[htbp]
\includegraphics[scale=0.5]{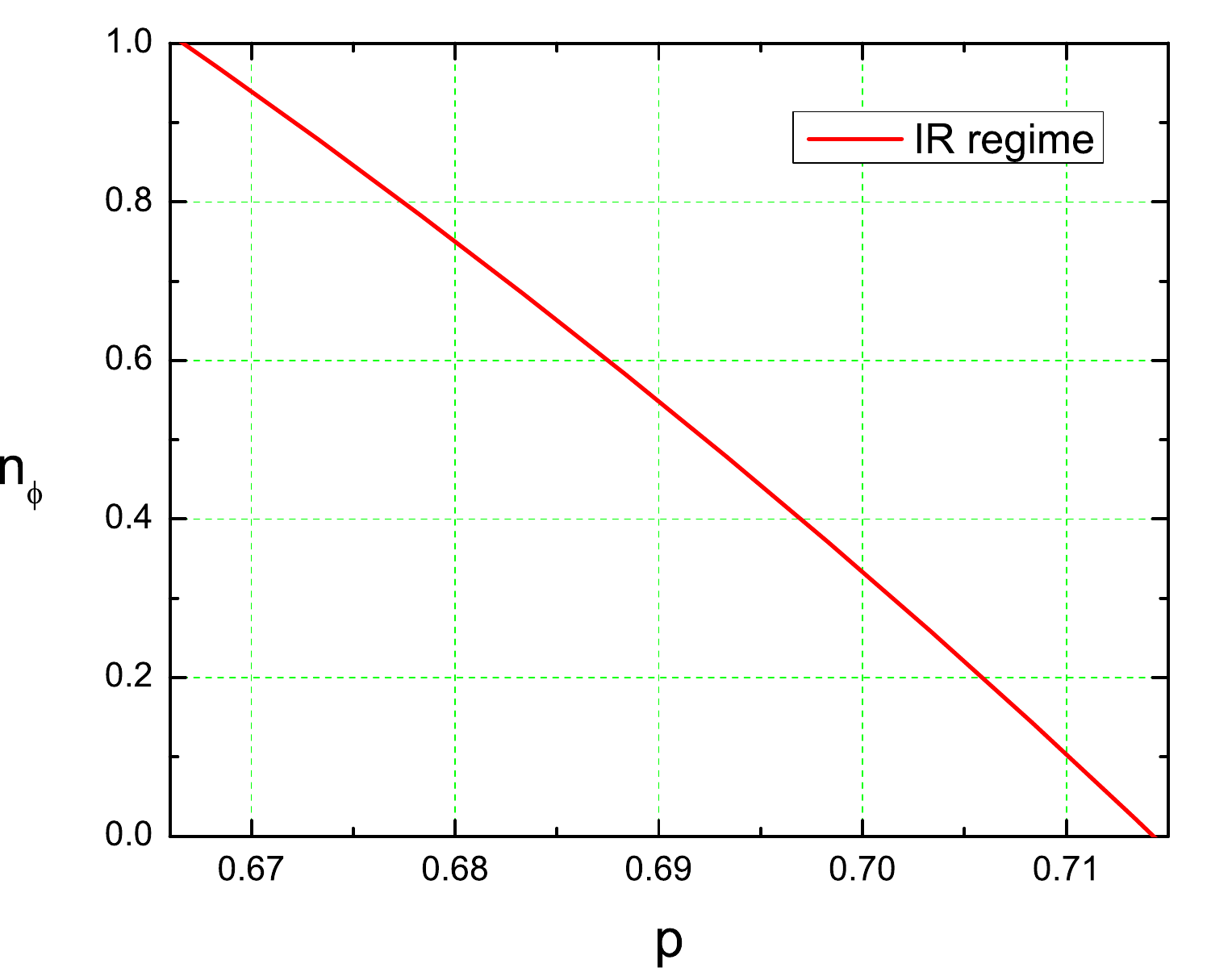}
\caption{The spectral index of the primordial perturbations with a standard dispersion relation as a function of the parameter $p$ in the IR regime.} \label{fig:nsnup}
\end{figure}
In Fig. \ref{fig:nsnup} we numerically plot the spectral index of the primordial perturbations under the effect of the $a''/a$ term. From the figure, one can see that the spectrum can also be red tilted when $p$ is above $2/3$. This implies that, when the universe experiences the contracting phase, one expects the background equation-of-state parameter to be slightly less than $w=0$ so that the primordial perturbations can form a power spectrum with a red tilt as observed by current CMB experiments. It is interesting to study what specific mechanisms can implement such a scenario in the future research.

\subsection{A brief summary}\label{subsec:nmbsummary}

So far we have reviewed the new matter bounce paradigm from both the theoretical setup and observational implications. As an alternative to inflationary cosmology, this paradigm is competitively successful in explaining the CMB observations. Moreover, this scenario has the following advantages that cannot be satisfied by inflation models. First of all, in the new matter bounce paradigm the big bang singularity is replaced by a nonsingular bouncing phase and thus the initial singularity problem is resolved. Secondly, since the power spectrum of cosmological perturbation was generated in the contracting phase during which the physical wavelengthes of these observable modes do not change dramatically, there is no Trans-Planckian problem for generic nonsingular bouncing cosmologies if the bounce scale is below the Planck scale. This profound property of cosmological perturbations is also in agreement with the theoretical starting point that one can describe the nonsingular bouncing cosmologies by the effective field approach without modifying Einstein's General Relativity.

Moreover, as shown in previous subsections, the (new) matter bounce paradigm usually predicts a large amplitude of primordial gravitational waves and hence the corresponding theoretical upper bound of the tensor-to-scalar ratio is of order unity. Accordingly, to apply the recent CMB primordial polarization measurement, we are able to constrain the parameter space of this type of models. Our analysis indicates that a reasonable matter-ekpyrotic bounce is expected to occur at an energy scale slightly less than the GUT scale and the duration of the bouncing phase is about ${\cal O}(10^4)$ Planck times.

At last, we discussed two possible mechanisms for generating a red tilt for the power spectrum of primordial perturbations in matter bounce cosmologies. One is to consider the UV modifications to the dispersion relation of primordial perturbation modes, and the other is to deform the background evolution by requiring the equation-of-state parameter to be slightly less than zero. Both mechanisms can successfully lead to a red tilted power spectrum. The difference is that the first one is only associated with the UV modes while the latter is generally associated with the IR modes. Although we have pointed out the possibility of realizing a red tilt in matter bounce cosmologies, it would be interesting to study the explicit realizations from theoretical model buildings. Furthermore, as has been noticed in \cite{Cai:2011zx}, there could exist different approaches for realizing a red tilted power spectrum if one takes into account the interaction terms among the matter fields in primordial epochs. These interactions are also associated with the reheating process of a universe in bouncing cosmologies \cite{Cai:2011ci}. These topics are expected to inspire more studies in the future.

\section{Toward the big bang in inflationary cosmology: the matter-bounce inflation scenario}\label{sec:bounce2}

In the previous section we studied the paradigm of the new matter bounce cosmology which can explain the CMB observations without inflation. However, there exists another interesting scenario that can solve the initial singularity problem within the context of inflationary cosmology, which suggests that the inflationary period is preceded by a nonsingular bounce and before that the universe has experienced a sufficiently long contracting phase with a pressureless dust matter being dominant. It is interesting to notice that the study of this paradigm is of observational interest as will be explained in the following subsection. Therefore, in this section, we would like to first review the latest observational constraint upon this scenario, and then go to the discussion of the model building.

\subsection{Observational constraints and cosmological implications}\label{subsec:mbi_constraint}

One interesting prediction of the matter-bounce inflation paradigm is that the amplitudes of the power spectra of primordial perturbations both of scalar and tensor types may undergo a jump feature at a critical scale. It was first pointed out in \cite{Liu:2010fm}, and recently generalized in \cite{Xia:2014tda}, that a series of parameterizations of primordial power spectra can be applied to confront with the latest cosmological data. In particular, we can perform a global fit and utilize the corresponding step feature to derive a novel method for constraining the bounce parameters.

To begin with, we briefly discuss the generation of primordial perturbations in the framework of a flat FRW Universe. The standard causal process suggests that, these observed large scale perturbations should initial emerge from very tiny spacetime fluctuations inside a Hubble radius at some primordial era, and then exit the Hubble radius to become classical perturbations and eventually re-enter at late times. This mechanism is robust to both the curvature perturbations and the relic gravitational waves. The dynamics of cosmological perturbations are convenient to be analyzed by tracking a Fourier mode $v_k$ along with the background evolution. In the context of General Relativity, if the background equation-of-state parameter is constant, then one can write down the equation of motion for the Fourier mode as follows,
\begin{eqnarray}\label{eom_mbi}
 v_k''+(k^2-\frac{a''}{a})v_k=0~,
\end{eqnarray}
where $a$ is the scale factor of the universe and the prime denotes the derivative with respect to the comoving time.

Specifically, the scale factor often evolves as $a(t)=a_B({t}/{t_B})^{1/\epsilon}$, where the subscript ``$B$" denotes any reference time which in our case can be referred as the bounce point. Note that the parameter $\epsilon$ is defined by Eq. \eqref{epsilon} in Section \ref{sec:inflation}. Its physical meaning is associated with the dynamics of background universe. For example, if the universe is in the inflationary stage as discussed in Section \ref{sec:inflation}, $\epsilon$ corresponds to the slow roll parameter; however, if the universe is in some other phase under a constant equation-of-state parameter $w$, then one has
\begin{eqnarray}
 \epsilon = \frac{3}{2}(1+w) ~.
\end{eqnarray}
Particularly, $\epsilon= 3/2$ for a dust matter dominated universe with $w=0$. Then, to transform into the comoving time, one gets
\begin{eqnarray}
 a(\eta) = a_B \left( \frac{\eta}{\eta_B} \right)^{1/(\epsilon-1)} ~,
\end{eqnarray}
and hence, the comoving Hubble parameter is given by
\begin{eqnarray}
 {\cal H} \equiv \frac{a'}{a} = \frac{1}{(\epsilon-1)\eta} ~.
\end{eqnarray}
From the above expression, one can immediately learn that, for inflation with $\epsilon\ll1$ there is $|{\cal H}| \simeq |1/\eta|$. Moreover, for a dust matter dominated universe with $\epsilon=3/2$, then $|{\cal H}| \simeq |2/\eta|$. Additionally, there is a very useful relation for the effective mass term given by,
\begin{eqnarray}\label{mass2}
 \frac{a''}{a} = \frac{(\mu^2-{1}/{4})}{\eta^2} ~, ~~~{\rm with}~~ \mu = \pm \frac{(\epsilon-3)}{(2\epsilon-2)} ~,
\end{eqnarray}
which can be derived from Eq. \eqref{mass}.

Following the method developed in \cite{Cai:2009hc} and applied in Sec. \ref{subsec:nmb_redtilt}, we study the asymptotic solutions to the perturbation equation in the sub-Planckian and super-Planckian regimes, respectively. First, we assume that all primordial perturbations originate from vacuum fluctuations inside the Hubble radius and hence, the WKB approximation suggests that
 $v_k^{i}\simeq \exp{(-i\int^\eta kd\tilde\eta)}/\sqrt{2k}$,
when $|k\eta|\gg1$. This is in agreement with the asymptotic solution of Eq. (\ref{eom_mbi}) when the last term ${a''}/{a}$ is negligible. Second, another asymptotic solution to Eq. (\ref{eom_mbi}) can be obtained at super-Hubble scales as
 $v_k \sim \eta^{{1}/{2}} [ c(k)\eta^{-|\mu|} ]$,
in the limit of $|k\eta|\ll1$. By matching these two solutions around the moment of Hubble crossing $|k\eta|\sim1$, we can obtain the super-Planckian perturbation mode as
\begin{eqnarray}\label{solution}
 v_k(\eta) \simeq \frac{1}{\sqrt{2k}} (k\eta)^{\frac{1}{2}-|\mu|}~.
\end{eqnarray}
The above solution is robust to primordial perturbations of both scalar and tensor types. However, the definitions of the power spectra of these two modes are different. Specifically, for primordial tensor modes, the power spectrum is defined as
\begin{eqnarray}
 P_T\equiv \frac{4k^3}{\pi^2}|\frac{v_k}{a}|^2 ~.
\end{eqnarray}
Moreover, for the scalar modes, the spectrum is expressed as
\begin{eqnarray}
 P_S \equiv \frac{k^3}{2\pi^2}|\frac{v_k}{z}|^2 ~,
\end{eqnarray}
with $z=a\sqrt{2\epsilon}$ being introduced.

Substituting the solution \eqref{solution} into these spectra, one can find that they could be nearly scale invariant either in the matter-dominated contracting phase or during inflationary stage. Accordingly, the power spectrum of primordial perturbations is overall roughly scale invariant in the matter-bounce inflation paradigm. However, as has been discussed previously, the comoving Hubble parameter evolves as $|{\cal H}| \simeq |2/\eta|$ during the matter contraction but becomes $|{\cal H}| \simeq |1/\eta|$ during inflation. If we require that the background evolution is smooth around the bounce point along the comoving time, then the comoving Hubble radius $ R_{\cal H} \equiv 1/|{\cal H}|$ would experience a rapid increase from $|\eta_B/2|$ to $|\eta_B|$. As a result, for the matter-bounce inflation model, the corresponding amplitude of primordial perturbations could achieve a jump feature around the scale $k_B$ comparable to the bounce scale. A much more detailed calculation yields that $P_T = {H^2}/{2\pi^2}$ when $k<k_B$ while $P_T = {2H^2}/{\pi^2}$ when $k\geq{k}_B$ for the model of matter-bounce inflation.

Following the method developed in \cite{Xia:2014tda}, we adopt the phenomenological parameterizations of primordial power spectra as follows,
\begin{align}\label{P_ST}
 P_{S} &= P_{m}+\frac{P_{\zeta,i}-P_{m}}{2} \left\{1+\tanh\big[T_B\log_{10} \frac{k}{k_B}\big]\right\}
 \nonumber \\
 P_{T} &= P_T^m+\frac{P_T^i-P_T^m}{2} \left\{1+\tanh\big[T_B\log_{10} \frac{k}{k_B}\big]\right\}
\end{align}
where we have introduced $P_T^m \equiv {H^2}/{2\pi^2}$ and $P_T^i \equiv {2H^2}/{\pi^2}$. In addition, $P_{\zeta,i}$ is the regular power spectrum during inflation which takes the form of \eqref{PS}, and $P_{m}$ is the scalar spectrum before the bounce which has to be less than $P_{\zeta,i}$. Correspondingly, $P_{\zeta,i}$ is characterized by two parameters, the amplitude $A_S$ and the spectral index $n_S$. Moreover, in order to determine the amplitude of primordial tensor modes, the standard process is to introduce the tensor-to-scalar ratio $r\equiv A_T/A_S$ as introduced in \eqref{ttsratio}.

Similar to this analysis, we would like to define a bounce-to-inflation ratio of the power spectrum as
\begin{eqnarray}
 r_B \equiv \frac{P_m}{A_S}~,
\end{eqnarray}
so that the amplitude of the scalar spectrum before the bounce can be characterized. As we will show later, this newly introduced parameter is very convenient to be constrained in numerical computation. In addition, from the parameterizations of primordial power spectra \eqref{P_ST}, one can see that there exist two more parameters, $k_B$ and $T_B$. Physically, $k_B$ is associated with the occurrence scale of the step feature in the spectra, and $T_B$ is determined by the duration of the bouncing phase and hence can depict the slope of this jump. As a result, the matter-bounce inflation scenario involves three more parameters and from their physical properties, it is apparent that they are highly correlated. For the convenience of numerical analyses, we can fix the value of $T_B$ as the best fit value derived from the combined Planck, WMAP and BICEP2 data, and then perform detailed computations to constrain the remaining two parameters, $k_B$ and $r_B$.

\begin{figure}[htbp]
\includegraphics[scale=0.4]{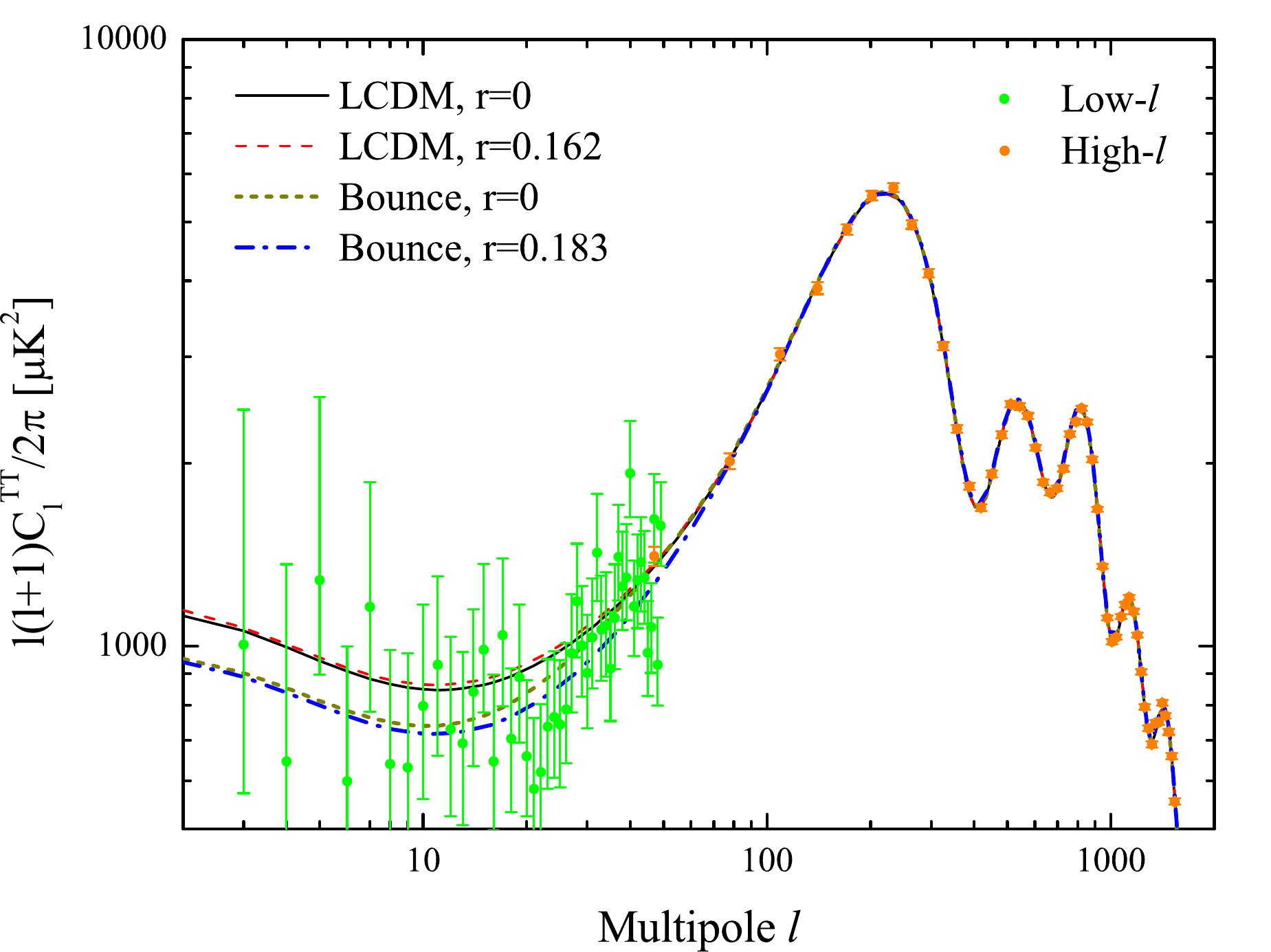}
\caption{ Signatures of the bounce effect on the CMB temperature power spectra for four early universe models. We compare the $\Lambda$CDM model and the matter-bounce inflation models with and without using the BICEP2 data. From \cite{Xia:2014tda}. }  \label{fig:bestCMB_TT}
\end{figure}

\begin{figure}[htbp]
\includegraphics[scale=0.4]{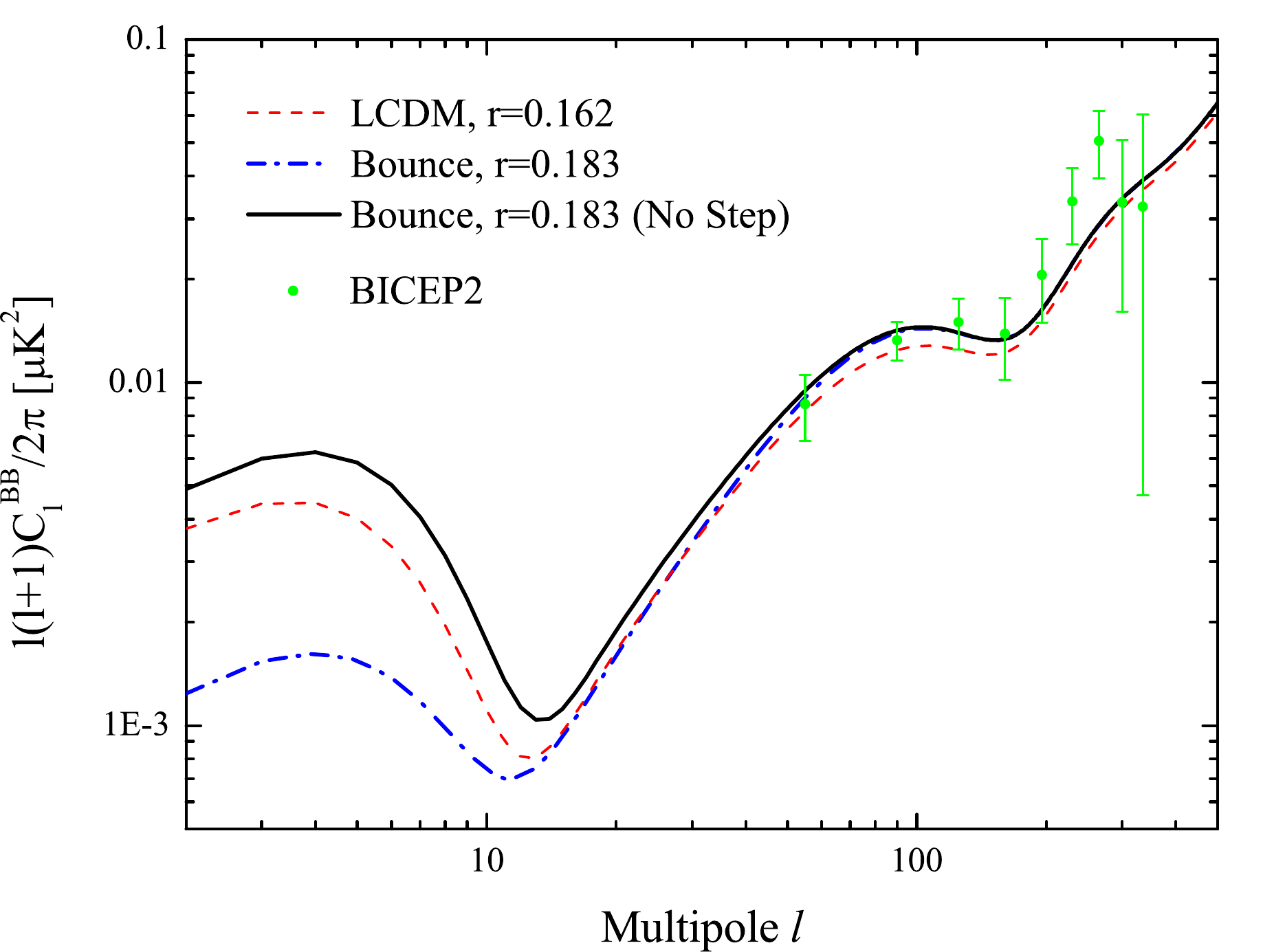}
\caption{ Signatures of the bounce effect on the CMB B polarization power spectra for four early universe models. We compare the $\Lambda$CDM model and the matter-bounce inflation models when using the BICEP2 data. From \cite{Xia:2014tda}. }  \label{fig:bestCMB_BB}
\end{figure}

From the expression in \eqref{P_ST}, one easily observes that, when $P_T^m=P_m=0$ and in the limit where $k_B\rightarrow0$, the spectra reduces to the regular result in standard inflationary paradigm. To test the potentially observable signals of the bounce in CMB measurements, we adopt $k_\mathrm{pivot} = 0.05$ Mpc$^{-1}$. Then we demonstrate the bounce effect on the CMB temperature and polarization spectra with different models in Figs. \ref{fig:bestCMB_TT} and \ref{fig:bestCMB_BB}. In particular, we consider the inflationary $\Lambda$CDM model and three matter-bounce inflation models with different parameter values as depicted in the figures. We can explicitly see that there is a suppression on both spectra at large angular scales.

As it is already known, the BICEP2 data strongly favor a nonzero primordial tensor spectrum, namely $ r = 0.2 ^{+0.07}_{-0.05}$ in the intermediate $\ell$ regime ($50 < \ell < 100$) \cite{Ade:2014xna}. This significantly large value of $r$, if it still exists in the low $\ell$ regime, can obviously lead to extra power in the CMB temperature spectrum and make it difficult to fit to the low $\ell$ Planck data. In the matter-bounce inflation models, however, the amplitudes of both scalar and tensor spectra are able to be suppressed in the low $\ell$ regime and therefore greatly relax this observational tension between the Planck and BICEP2 data sets. In this regard, we can claim that the matter-bounce inflation paradigm can explain the combined CMB observations very well compared to the $\Lambda$CDM model.

Since the BICEP2 experiment can only measure the B-mode power spectrum at scales $\ell > 30$ where the suppression effect is small, the median value of the tensor-to-scalar ratio $r$ in the matter-bounce inflation paradigm is close to that obtained in the standard $\Lambda$CDM model. Moreover, the suppression effect becomes very significant at very large scales $\ell < 20$ which is expected to be sensitive to the forthcoming CMB polarization data as will be released soon by the Planck team. Therefore, it is very promising to further constrain or even probe the observational signals of a nonsingular bounce in near future.

In Ref. \cite{Xia:2014tda}, we also performed a global fit on various model parameters in this paradigm based on a generalized CosmoMC code package. We refer to that paper for details, and briefly summarize the main results as follows. According to the global fitting analysis, the numerical constraint on the model parameters from the combined Planck, WMAP and BICEP2 data are found to be the following: the bounce parameter scale $\log_{10}(k_{\rm B}/{\rm Mpc}^{-1})=-2.4\pm0.2$ and the bounce-to-inflation ratio $r_B\equiv P_{\rm m} / A_{\rm s} = 0.71\pm0.09$, at $68\%$ C.L., respectively. Note that, the suppression effect of the CMB temperature anisotropies may also be achieved in some string inspired inflation models \cite{Ashoorioon:2008qr}.

\subsection{A toy model by means of the Horndeski and ghost-condensate operators}\label{subsec:mbi}

Qualitatively speaking, a matter-dominated contracting universe is very similar to what our universe has experienced since the moment of radiation-matter equality until dark energy dominates over, except that the sign of their Hubble parameters is opposite. However, it is not trivial to achieve such an extended inflationary model that can stably realize a bouncing phase before inflation. One ought to be aware of the following theoretical constraints.

First of all, it is not trivial to build a matter-bounce inflation model that is stable enough against any ghost mode since this scenario involves a violation of NEC which is required by the bouncing phase. Secondly, one should be aware of potentially dangerous gradient instabilities that might bring an unexpected growth to primordial perturbations. Any specific model that realizes the matter-bounce inflationary scenario must have a canonical Lagrangian with all higher order operators being suppressed in the IR regime and therefore, the model is free from the graceful-exit problem.

To stably violate the weak energy condition is of theoretical interest in various models of very early universe physics. One plausible mechanism of achieving such a scenario is to make use of a ghost condensate field in which the kinetic term for the inflaton takes a non-vanishing expectation value in the infrared regime \cite{ArkaniHamed:2003uy, ArkaniHamed:2003uz}. However, this type of model often suffers from gradient instabilities when the universe exits from the inflationary phase to the normal thermal expansion phase.

Another approach to realize the weak energy condition violation is to make use of a Galileon type field \cite{Nicolis:2008in, Deffayet:2011gz, Horndeski:1974wa}. The key feature of this model is that it contains higher order derivative terms in the Lagrangian while the equation of motion remains second order, and thus does not necessarily lead to the appearance of ghost modes. Recently the Galileon-type field has been applied to interpret the late time acceleration of the universe in \cite{Deffayet:2010qz}. It was also applied to drive the inflationary phase at early times, for instance, see \cite{Kobayashi:2010cm, Kamada:2010qe}, and see \cite{Ohashi:2012wf} for the discussion of the stability issue. The Galileon model might realize a nonsingular bounce as well, but was found to suffer from a severe graceful-exit problem from the bouncing phase to the phase of regular thermal expansion \cite{Qiu:2011cy, Easson:2011zy}.

Inspired by the success of the model building in the new matter bounce paradigm \cite{Cai:2012va}, we suggest in this section that the matter-bounce inflation scenario could be achieved by a single scalar field possessing the desirable features of both the ghost-condensate and Galileon-inspired models. In particular, we phenomenologically propose a Lagrangian of the following form
\begin{eqnarray} \label{L_MBI}
 {\cal L} = X -V(\phi) -\tilde{g}(\phi) M_p^2 (2X)^{\frac{1}{2}} + \frac{\tilde\gamma(\phi)}{M_p} (2X)^{\frac{1}{2}} \Box\phi ~,
\end{eqnarray}
which is similar to the form introduced in Eq. \eqref{L_KGB}. The main differences are as follows. For one thing, the coefficient of the Horndeski operator is generalized as a function of $\phi$; for another, we have modified the ghost-condensate-like operator by introducing a $X^{1/2}$ term.

Note that, in the expression of \eqref{L_MBI}, the first two terms are the canonical kinetic term and the potential for the inflaton field and thus are of canonical forms. The third term is introduced to realize a stable ghost condensate in the simplest way. Note that this term was earlier applied to explain the late time acceleration of the universe and was dubbed as the ``cuscuton" term \cite{Afshordi:2006ad, Afshordi:2007yx}. In particular, when $g>0$ the ghost condensate state would form and in the opposite case the scalar field remains the canonical form. The Horndeski operator $G$ is introduced to stabilize the propagation of cosmological perturbations when the bouncing phase occurs. Its effect is automatically suppressed at low energy scales when $X$ becomes small. Consequently, the model under consideration could satisfy the theoretical limits pointed out at the beginning of this section. The research along this line is particularly interesting from a phenomenological perspective and we would like to leave it for future studies.

\section{Conclusion}\label{sec:conclusion}

Since the 1990s, we witnessed the birth of the era of precision cosmology. Starting from the COBE satellite, we have been able to measure the CMB map with high precision by virtue of accumulated observational data from the space telescopes such as the WMAP and Planck spacecrafts; meanwhile, the ground observatories like the SPT and BICEP2 collaborations have made significant developments on signaling the CMB at specific angular scales. With these great improvements of cosmological observations, we have arrived at the urgent status of addressing the theoretical question of how the current early universe picture can arise from fundamental theories. In particular, the recent announcement of detection of non-vanishing primordial B-mode polarization by the BICEP2 collaboration has led the interest of cosmologists into the accurate examinations of very early universe models and inspired a number of extensive studies, for instance, see \cite{Kehagias:2014wza, Brandenberger:2014faa, Li:2014cka, Cai:2014xxa, Cai:2014bda, Xia:2014tda, Lizarraga:2014eaa, Zhao:2014rna, Ma:2014vua, Harigaya:2014qza, Gong:2014qga, Miranda:2014wga, Hertzberg:2014aha, Gerbino:2014eqa, Wang:2014kqa, Moss:2014cra, Bonvin:2014xia, Zhang:2014dxk, Lyth:2014yya, DiBari:2014oja, Hazra:2014jka, Smith:2014kka, Hossain:2014coa, Chung:2014woa, Hu:2014aua, Cai:2014hja} and references therein.

So far the standard inflationary $\Lambda$CDM paradigm has made a number of initial achievements in the past decade, in particular addressing theoretical issues of the hot big bang cosmology. Its prediction of nearly scale invariant primordial power spectra has been gradually verified by several CMB experiments as reviewed in the article. This scenario, however, suffers from a criticism that at the moment of the big bang, there is a singularity of zero volume and infinite energy. The corresponding physics is beyond all knowledge we have today, and thus we expect to find a resolution to this problem.

Most of the recent theoretical developments in quantum gravity, such as string theory and loop quantum gravity, strongly indicate that there exists a fundamental minimal length scale in the universe. Based on this profound quantum property, the nonsingular bouncing cosmology was proposed as a promising paradigm attempting to avoid the initial singularity. This cosmological paradigm suggests that there was a contracting phase preceding the regular thermal expansion of the universe, and in the contraction the universe was dominated by a pressureless matter component, the corresponding power spectra of primordial perturbations can be almost scale-invariant similar to the inflationary predictions. Accordingly, this so-called matter bounce scenario could explain the CMB observations and the formation of the LSS just as successfully as inflationary cosmology, but was found to suffer from the BKL instability against the backreaction of primordial anisotropies.

Along this line, a so-called new matter bounce (or matter-ekpyrotic bounce) cosmology was recently proposed to combine the advantages of the matter bounce scenario and the ekpyrotic cosmology which can dilute unwanted primordial anisotropies. Moreover, this new paradigm can be described within the context of General Relativity by using techniques of effective field approach which can efficiently approximate the unknown underlying theory. This approach is widely applied in general topics of physics. By virtue of this approach, we are able to construct the models of the new matter bounce paradigm in which the universe starts its evolution from a contracting phase in the far past and evolves without pathologies. In Sec. \ref{sec:bounce1} we studied in detail the specific model building of this paradigm by including the ghost condensate and Horndeski operators into the field Lagrangian. Moreover, we have confronted this type of model with the latest cosmological observations and shown that a new matter bounce model with a bounce scale being lower than the GUT scale and a sufficiently long bouncing phase is favored by the present data.

Furthermore, in light of the nonsingular bouncing cosmology, we discussed a generalized bounce inflationary cosmology with a matter-dominated contraction preceding an inflationary phase. This scenario has been obtained by the effective field approach and also in loop quantum cosmology. In particular, this paradigm is of significant observational interest due to its predictions on the featured power spectra of both scalar and tensor perturbations. Due to this notable phenomenological feature, we analyzed in Sec. \ref{subsec:mbi_constraint} the latest observational constraint on the parameter space of this paradigm from the combined Planck and BICEP2 data. This feature is very promising to be confirmed (or ruled out) by the Planck polarization data which will be released in the near future. Afterwards, we attempted to explore the possibility of realizing this scenario by a well-behaved theoretical model. Though it is yet to be finalized, we can explicitly see that the associated model building deserves a comprehensive analysis in the future.

Concerning the theoretical implications of nonsingular bouncing cosmologies, an important lesson is to break a series of singularity theorems proven by Penrose \cite{Penrose:1964wq} and Hawking \cite{Hawking:1969sw} about half a century ago and later developed into a cosmological framework by Borde, Vilenkin and Guth \cite{Borde:2001nh}. Such an issue is often accompanied with the violation of NEC if we restrict our study within the framework of classical General Relativity. As a result, the phenomenologies of a nonsingular bouncing cosmology in the very early universe could be associated with the quintom scenario as inspired by the dark energy study of the late time acceleration as reviewed in \cite{Cai:2009zp}. Along this line, we expect that there might be a bridge connecting the physics of very early universe with that at late times.

Eventually, we would like to end this article by commenting on some unsettled issues of bouncing cosmologies. Although the topic of nonsingular bouncing cosmologies has become exciting and attracted more and more cosmologists' attention in recent years (for example, see \cite{Lilley:2011ag, Graham:2011nb, Linsefors:2012et, Linsefors:2013cd, Osipov:2013ssa, Qiu:2013eoa, Li:2013hga, Xue:2013bva, Garriga:2013cix, Bamba:2013fha, Bamba:2014mya, Li:2014era, Gao:2014hea, Forte:2014yia, Li:2014qwa} for recent interesting studies), the detailed paradigm is still far from being specified. Namely, in the present article we have analyzed two plausible bouncing cosmologies that can satisfy cosmological observations equally well. In the following we present the main questions, which in our opinion are crucial to the development of bouncing cosmologies in the future.
\begin{itemize}
\item Firstly, the fundamental theory that gives rise to a nonsingular bounce is still not understood. The connection to quantum physics is unresolved.
\item Secondly, the potentially observable signatures of a nonsingular bounce need to be verified. To distinguish them from other early universe paradigms needs to be addressed.
\item Lastly, if the universe did experience a bounce, this may necessitate another bounce in the future. It may be possible that it is cyclical nature.
\end{itemize}

For a very long time, physicists have been continuously making attempts to address the above questions. However, we believe that the final resolutions will eventually be discovered under the combined efforts of both observational and theoretical cosmology. If a nonsingular bounce is verified, the knowledge about our universe will usher in a new era.

\section*{Acknowledgments}

%{\it Acknowledgments.---}
I am indebted to Robert Brandenberger and all collaborators for long-term collaborations in the field of theoretical cosmology. I am also grateful to Damien Easson, Hong Li, Jerome Quintin, Emmanuel Saridakis, Edward Wilson-Ewing, Jun-Qing Xia for kind permissions to include figures from previous works. In addition, I acknowledge to Prof. Yipeng Jing and Prof. Xinmin Zhang for the invitation to contribute to the special issue of ``Science China: Physics, Mechanics \& Astronomy". This work is supported in part by NSERC and by the Department of Physics at McGill.

\end{document}